\documentclass{jfm}
\usepackage{graphicx,xcolor}
\usepackage{epstopdf,epsfig}

\shorttitle{The dynamics of stratified horizontal shear flows at low P\'{e}clet number}
\shortauthor{L. Cope, P. Garaud and C. P. Caulfield}

\title{The dynamics of stratified horizontal shear flows at low P\'{e}clet number}
\author{Laura Cope\aff{1}
  \corresp{\email{lauracope@cantab.net}},
  P. Garaud\aff{2}
 \and C. P. Caulfield\aff{3} \aff{1}}

\affiliation{\aff{1}Department of Applied Mathematics and Theoretical Physics, University of Cambridge, Centre for Mathematical Sciences, Wilberforce Road, Cambridge CB3 0WA, UK
\aff{2} Department of Applied Mathematics, Baskin School of Engineering, University of California Santa Cruz, Santa Cruz, CA 95064, USA
\aff{3} BP Institute, University of Cambridge, Madingley Rise, Madingley Road, Cambridge CB3 0EZ, UK}

\begin{document}

\maketitle

\begin{abstract}
We consider the dynamics of a vertically stratified, horizontally-forced Kolmogorov flow.  Motivated by astrophysical systems where the Prandtl number is often asymptotically small, our focus is the little-studied limit of high Reynolds number but low P\'{e}clet number (which is defined to be the product of the Reynolds number and the Prandtl number). Through a linear stability analysis, we demonstrate that the stability of two-dimensional modes to infinitesimal perturbations is independent of the stratification, whilst three-dimensional modes are always unstable in the limit of strong stratification and strong thermal diffusion. The subsequent nonlinear evolution and transition to turbulence is studied numerically using direct numerical simulations. For sufficiently large Reynolds numbers, four distinct dynamical regimes naturally emerge, depending upon the strength of the background stratification. By considering dominant balances in the governing equations, we derive scaling laws for each regime which explain the numerical data.
\end{abstract}

\begin{keywords}
Authors should not enter keywords on the manuscript, as these must be chosen by the author during the online submission process and will then be added during the typesetting process (see http://journals.cambridge.org/data/\linebreak[3]relatedlink/jfm-\linebreak[3]keywords.pdf for the full list)
\end{keywords}


\section{Introduction}

Statically stable stratified flows, where the background equilibrium  fluid density decreases (at least on average) upwards in a gravitational field, are ubiquitous. Examples in geophysics include atmospheres, oceans and lakes, while they also occur on astrophysical scales in planetary and stellar interiors. A key physical process in such flows is that fluid parcels perturbed vertically from their equilibrium position experience a restoring `buoyancy force'. Furthermore, it is generic that the fluid will also have a spatially and temporally varying background velocity distribution that is expected to interact with the background `stable' stratification. 

In many cases, the flow becomes turbulent, and the interaction between the turbulence and the stratification is a major source of vertical transport in geophysical flows \citep{Ivey2008,Ferrari2009} and stellar interiors \citep{Zahn1974,Zahn92,SpiegelZahn92}. What has become known as `stratified turbulence' in the geophysical literature exhibits a wide range of dynamical, and often counter-intuitive behaviours, not least because it leads to complex, and still controversial, irreversible energetic exchange pathways between the kinetic, potential and internal energy reservoirs. Understanding and modelling those pathways, in particular the `efficiency' of mixing (associated with the irreversible conversion of kinetic energy into potential energy) is of great importance for larger-scale descriptions of geophysical flows, such as weather forecasts, ocean circulation simulation or indeed climate models, and astrophysical flows that regulate planetary and stellar evolution. In what follows, we first describe the current understanding of stratified turbulence in geophysical flows, and explain why these results need to be revisited in the astrophysical context, which is the purpose of this work.

\subsection{Stratified turbulence in geophysical flows}
\label{sec:intro1}

There has been a great amount of research into transition, turbulence and mixing in stratified flows with focus on the relevance to atmospheric and oceanic flows \citep[e.g.][]{ Ivey2008}. Within this context, the simplest idealised (yet commonly considered) situation has three fundamental modelling assumptions: that the fluid velocity is solenoidal, i.e. 
$\nabla \cdot \mathbf{u} =0$; that the density differences within the flow are sufficiently
small for the `Boussinesq approximation' with a linear equation of state to be an appropriate model; and that the density variations in the fluid are associated with a single stratifying agent, avoiding the occurrence of `double diffusive' phenomena \citep[which may still be very important in a variety of different circumstances, see for example the reviews of][]{Schmitt1994,radko2013double,Garaud2018}. Without loss of generality, the density field $\rho$ may be assumed to be a function of temperature $T$ alone, such that
\begin{equation}
  \frac{(\rho-\rho_0)}{\rho_0} = - \alpha (T-T_0), \label{eq:eqstate}
  \end{equation} 
where $\rho_0$ and $T_0$ are reference densities and temperatures, and $\alpha$ is the thermal expansion coefficient.
Since temperature satisfies an advection-diffusion equation
\begin{equation}
	\frac{\partial  T }
	{\partial t}+ \mathbf{u} \cdot \nabla T = \kappa \nabla^2 T ,
	\label{eq:advdiff}
\end{equation}
where $\kappa$ is the thermal diffusivity, the density fluctuations also satisfy the same advection-diffusion equation.

Both irreversible mixing and turbulent viscous dissipation, leading respectively to irreversible changes in the potential energy and internal energy of the flow, rely inherently on the action of diffusive processes. Under the three simplifying assumptions above, the stratified fluid under consideration has only two relevant diffusivities: the kinematic viscosity $\nu$ quantifying the diffusivity of momentum; and $\kappa$, quantifying the diffusivity of density. 
Together with these diffusivities, there are at least three additional dimensional parameters required to describe a stratified flow: a characteristic velocity scale $U_c$, a characteristic length scale $L_c$, and a characteristic buoyancy frequency $N_c$ associated with the background buoyancy frequency profile $N_b(z)$, defined as 
\begin{equation}
	N_b^2(z) \equiv -\frac{g}{\rho_0} \frac{d \rho_b }{dz} = \alpha g \frac{d T_b}{dz} ,\label{eq:ncdef}
\end{equation}
where $g$ is the acceleration due to gravity, and $\rho_b$ and $T_b$ are background profiles of density and temperature, respectively. Note that the Boussinesq approximation requires that the total variation of a background scalar quantity $q_b(z)$ satisfies $L_c |d q_b/dz |\ll q_0$.

A natural set of non-dimensional parameters can be constructed as: a  Reynolds number $Re$ quantifying the relative magnitude of inertia to momentum diffusion by viscosity; a P\'eclet number $Pe$ quantifying the relative magnitude of inertia to the diffusion of the density; and a Froude number $Fr$ quantifying the relative magnitude of the inertia to the stratification, defined as
\begin{equation} \label{eq:parameter_definitions}
	Re \equiv \frac{U_c L_c}{\nu}, \quad Pe \equiv \frac{U_c L_c}{\kappa}= Pr Re, \quad \mbox{ and } \quad  Fr \equiv
	\frac{U_c}{N_c L _c} , \label{eq:rec}
\end{equation} 
where $Pr=\nu/\kappa$ is the Prandtl number. Note that for vertically sheared flows, the Froude number is related to a bulk Richardson number as 
\begin{equation}
Ri = \frac{N_c^2 L_c^2}{U_c^2} = Fr^{-2}.
\end{equation}
Also note that we have implicitly restricted our focus to a regime where the scales of motion are sufficiently small and fast so that the effects of rotation can be ignored, otherwise an additional parameter is necessary. 

As discussed in detail in \cite{Riley2000} and \cite{Brethouwer2007}, oceanic and atmospheric flows are often very strongly stratified, in the specific sense that if both  $L_c$ and  $U_c$ are identified  with typical scales of horizontal motions, then $Fr \ll 1$ ($Ri \gg 1$). Nevertheless, turbulence still occurs, at least in spatio-temporally varying patches \citep{Portwood2016}. This has profound  implications for understanding the dynamics of such flows.


\cite{Brethouwer2007}, following  \cite{BillantChomaz2001}, demonstrated that when both $Re \gg 1 $ and $Fr \ll 1$,
several different flow regimes are possible. Each regime can be understood as a distinct dominant balance between various terms in the Navier-Stokes equations, dependent on their relative sizes. Of central significance to these balances, however, 
are two additional geophysically-motivated parameter choices, both of which we wish to revisit in this manuscript which aims to extend this work to astrophysically-relevant flows. The first of these parameter choices is motivated by the expectation (and empirical observation) that `strong' stratification leads to anisotropy in the flow, so the characteristic vertical length scales $L_v$ are expected to be very different from characteristic horizontal length scales $L_h \equiv L_c$.
The second relies on the fact that the Prandtl number is of order unity or larger in geophysical flows. Typically, $Pr \sim O(1)$ for gases (e.g. $Pr \simeq 0.7$ for air) while for fresh water $Pr \sim O(10)$ (with some variability with temperature and pressure, although the canonical value is chosen to be $7$). If the density variations are due to salinity with diffusivity $D$ rather than temperature differences, the analogous ratio of diffusivities, known as the Schmidt number $Sc=\nu/D\sim O(1000)$, is even higher.

With these two further choices, \cite{Brethouwer2007} discussed three particular regimes which are worthy of comment.
The first, originally considered by \cite{Lilly1983} \citep[also see][for further discussion]{Riley2000} has  $Re \gg 1$ and $Fr \ll 1$, yet $L_v/L_c \gg Fr$ and also $L_v/L_c \gg 1/\sqrt{Re}$. With these scalings, all terms involving vertical derivatives (specifically diffusive terms and advective terms involving vertical velocity) are insignificant in the Navier-Stokes equations, and so the governing equations reduce to the evolution equations for an incompressible and inviscid `two-dimensional'
horizontal velocity ${\mathbf u}_h(x,y,t)$. Furthermore, since $Pr \gtrsim O(1)$, diffusive terms in the density equation can also be ignored, and quasi-two dimensional (though possibly layerwise) flow evolution is expected.

The other two regimes discussed in detail by \cite{Brethouwer2007} still rely essentially on the fact that $Pr \gtrsim O(1)$. They  also exploit the insight of \cite{BillantChomaz2001} that the vertical length scale should not be externally imposed, but should emerge as a property of the flow dynamics. In that respect, as presented in detail by \cite{Brethouwer2007}, a key parameter is the quantity commonly referred to as the `buoyancy Reynolds number' $Re_b$, defined as
\begin{equation}
	Re_b \equiv Re Fr^2 = \frac{U_c^3}{\nu L_c N_c^2} . \label{eq:rebdef}
\end{equation}
When $Re_b \lesssim O(1)$, (but still with $Pr \gtrsim O(1)$), a viscously affected regime is expected, where horizontal advection is balanced by viscous diffusion, specifically associated with vertical shearing. This regime, much more likely to be relevant in experiments (or simulations) rather than in geophysical applications, has $L_v/L_c \sim 1/\sqrt{Re}$, and does not exhibit a conventional turbulent cascade, but rather exhibits the effects of viscosity (and density diffusion) even at large horizontal scales.

Conversely, \cite{Brethouwer2007} showed that when $Re_b \gg 1$, viscous effects are insignificant (as is density diffusion since $Pr \gtrsim O(1)$) and the remaining terms (including the advection by the vertical velocity) become self-similar with respect to $z N_c/U_c$, with $z$ being the vertical coordinate aligned with gravity. This suggests strongly that $L_v \sim U_c/N_c$, or equivalently that the Froude number based on the vertical scale $L_v$, defined as 
\begin{equation}
	Fr_v \equiv \frac{U_c}{L_v N_c},  \label{eq:frv}
\end{equation}
should be of order one, so $L_v \ll L_c$. Such a vertical layer scale has been commonly observed in a wide variety of sufficiently high Reynolds number stratified flows \citep[e.g][]{Park1994,Holford1999,BillantChomaz2000,Godeferd2003,Brethouwer2007,Oglethorpe2013,Lucas2017,Zhou2019} and appears to be a generic property of high $Re_b$ and high $Pr$ stratified turbulence. This regime is characterised not only by anisotropic length scales but also by anisotropy in the velocity field, and hence the associated turbulence, leading \cite{Falder2016} to refer to this flow regime as the `layered anisotropic stratified turbulence' (LAST) regime. 

The vertical layering on the scale $L_v$ is key to understanding how turbulence can be maintained in the LAST regime despite the fact that $Fr \ll 1$. Indeed, these `layers' in the density distribution consist of relatively weakly stratified wider regions separated by relatively thinner `interfaces' with substantially enhanced density gradient. As such, local values of the buoyancy frequency can vary widely from the characteristic value $N_c$. When the local vertical shear is sufficiently strong compared to the local density gradient, then the gradient Richardson number $Ri_g$, defined as
\begin{equation}
	Ri_g \equiv \frac{- g}{\rho_0} \frac{\partial \rho / \partial z }{ \left|  \partial {\bf u}_h / \partial z \right|^2 },
	\label{eq:rig}
\end{equation}
can drop to values low enough for shear instabilities to be able to develop. If in addition the Reynolds number is sufficiently
large for inertial effects to be dominant, this allows for the possibility of turbulence  through the breakdown of shear instabilities, albeit with both spatial and temporal intermittency. 

It is crucial to appreciate that this LAST regime relies inherently on the assumption that $Pr \gtrsim O(1)$, as high Reynolds number thus implies high P\'eclet number, so localized turbulent events can erode the stratification and in turn participate in the formation or maintenance of the layers. Although appropriate for the atmosphere and the ocean, this fundamental assumption most definitely does not apply in the astrophysical context, where $Pr \ll 1$ (see below). As we now demonstrate, density layering is prohibited in that case, suggesting that LAST dynamics cannot occur. This raises the interesting question of whether analogous or fundamentally different regimes exist when $Pr \ll 1$.

\subsection{Stratified shear instabilities in stars}
\label{sec:intro2}

The fluid from which stars and gaseous planets are made is a plasma comprised of photons, ions, and free electrons. As a result, one of the main differences between astrophysical and geophysical flows is the value of the Prandtl number, which is much smaller than one as mentioned above. In typical stellar radiative zones, for instance, $\Pran$ usually ranges between $10^{-9}$ and $10^{-5}$ \citep[see][figure 7]{Garaudal15b}. The microphysical explanation for this difference is that heat can be transported by photons efficiently while momentum transport usually requires collisions between ions (which comprise most of the mass), so $\nu \ll \kappa$ and $\Pran \ll 1$. This crucially introduces the possibility of a new regime of flow dynamics where $Re \gg 1$ while $Pe = \Pran Re \ll 1$, which is never realized in geophysics. In fact, that possibility is always realized provided that the characteristic scale $L_c$ considered in (\ref{eq:parameter_definitions}) is sufficiently small.

Astrophysical fluids are also not incompressible. However, under a set of assumptions that are almost always satisfied sufficiently far below the surface of stars and gaseous planets, the Spiegel-Veronis-Boussinesq approximation \citep{SpiegelVeronis1960} can be used to reduce the governing equations to a form that is almost equivalent to that used for geophysical flows. In particular, $\nabla \cdot {\bf u} \simeq 0$, $(\rho-\rho_0)/\rho_0 \simeq - \alpha (T-T_0)$, and the temperature equation (\ref{eq:advdiff}) becomes
\begin{equation}
\frac{\partial T}{\partial t } + {\bf u} \cdot \nabla T  + w \frac{g}{c_p} = \kappa \nabla^2 T ,\label{eq:advdiffastro}
\end{equation}
where $c_p$ is the specific heat at constant pressure. In comparison with (\ref{eq:advdiff}), the new term $w g/c_p$ accounts for compressional heating (or cooling) as the parcel of fluid contracts (or  expands) to adjust to the ambient pressure as it moves downwards (or  upwards)  in a gravitational field. As a result, the background buoyancy frequency profile $N_b(z)$ is modified from (\ref{eq:ncdef}) to 
\begin{equation}
N_b^2(z) \equiv \alpha g \left( \frac{d T_b }{dz} + \frac{g}{c_p} \right),\label{eq:ncstardef}
\end{equation}
from which a new characteristic buoyancy frequency $N_c$ can be defined.

Interest in stratified shear instabilities at low Prandtl number and/or low P\'eclet number in stars dates back to \cite{Zahn1974}. In this regime, thermal dissipation greatly mitigates and modifies the effect of stratification in comparison
to flows with $Pr \gtrsim O(1)$. In particular, as demonstrated by \cite{Lignieres1999} \citep[see also][]{Spiegel1962,Thual1992}, a dominant balance emerges in the temperature equation whereby 
\begin{equation}
 w \left( \frac{d T_b }{dz} + \frac{g}{c_p} \right) \simeq \kappa \nabla^2 T, 
 \label{eq:lpeq}
\end{equation}
(at least to leading order in $Pe^{-1}$), showing that temperature fluctuations and vertical velocity fluctuations are slaved to one another (see more on this in section \ref{sec:model}). Mass conservation, combined with appropriate boundary conditions, then generally implies that the horizontal average of $T$ should be zero. Physically, this simply states that due to the very rapid diffusion of the temperature fluctuations (and hence density), perturbations cannot modify the background. Density layering is therefore prohibited, as stated above, so the local buoyancy frequency remains close to the background value $N_b(z)$ everywhere.

Another important consequence of this highly diffusive limit \citep{Lignieres1999} is that the P\'eclet and Froude (or Richardson) numbers are no longer independent control parameters for the system dynamics, but always appear together as $Pe/Fr^2$ or $Ri Pe$. \cite{Zahn1974} argued that, as a result, the threshold for vertical shear instability should be $Ri Pe  \lesssim  Re / Re_c$ where $Re_c$ is the critical Reynolds number for instability in unstratified, unbounded shear flows (which he estimated would be $O(1000)$). Zahn's criterion for instability is now commonly written as $Ri \Pran \lesssim K_Z$, where $K_Z  \sim O(10^{-3})$. 
This was recently independently confirmed using direct numerical simulations (DNSs) by \cite{Pratal2016} \citep[see also][]{PratLignieres13,PratLignieres14} and \cite{Garaudal17}, who found that $K_Z \simeq 0.007$. With the aforementioned estimates for $\Pran$, we see that shear-induced turbulence in low $Pe$ vertical shear flows is therefore likely for $Ri$ up to $\sim 10^2$ or higher. On the other hand, for astrophysical flows with $Ri \Pran \gg K_Z$, or for horizontally-sheared flows (see below), one may naturally ask whether any pathway to turbulence exists, since the density layering that is central to the LAST regime is not possible here. This paper aims to answer this question for the case of horizontally-sheared flows.

Before proceeding, however, it is useful to briefly review the most commonly used model of shear-induced mixing in stars \citep[see][for a more comprehensive review of the topic]{Lignieres2018}. \cite{Zahn92} considered successively both vertically-sheared flows and horizontally-sheared flows. For a vertically-sheared flow with characteristic shearing rate $S_c$, he argued based on work by \cite{Townsend58} and \cite{Dudis1974} that the largest unstable vertical scale in the flow would satisfy $Ri Pe_l \sim O(1)$, where here $Ri = N_c^2 / S_c^2$ and where $Pe_l \equiv S_c l^2 / \kappa$ is an eddy-scale P\'eclet number. This defines the characteristic Zahn scale $l_Z$ as
\begin{equation}
Ri \frac{S_c l_Z^2}{\kappa} \sim O(1) \quad \Rightarrow  \quad
l_Z \sim \sqrt{\frac{ \kappa}{Ri S_c} } \sim \sqrt{\frac{ \kappa S_c}{N_c^2} }.
\label{eq:Zahnscale}
\end{equation}
Using dimensional analysis, \cite{Zahn92} then proposed a simple expression for a turbulent diffusion coefficient, namely 
\begin{equation}
D_{\rm turb} \sim S_c l_Z^2  \sim  \frac{\kappa}{Ri}.
\label{eq:Dturb}
\end{equation}
The relevance of the Zahn scale to the dynamics of {\it low P\'eclet} number stratified turbulence in vertically-sheared flows was confirmed by \cite{Garaudal17} using direct numerical simulations (DNSs). They also verified that (\ref{eq:Dturb}) applies for flows that have both low P\'eclet number and sufficiently high Reynolds number, as long as $l_Z$ is much smaller than the domain scale, and $Ri \Pran \lesssim K_Z$ \citep[see also][]{PratLignieres13,PratLignieres14,Pratal2016}.

In the horizontally-sheared case, \cite{Zahn92} postulated \citep[following an argument attributed to][]{SchatzmanBaglin1991}, that while the turbulence would be mostly two-dimensional on the large scales owing to the strong stratification, it could become three-dimensional below a scale $L_c$ where thermal dissipation becomes important. This scale is by definition the Zahn scale, and is therefore given by (\ref{eq:Zahnscale}) where here $S_c = U_c / L_c$ (and $U_c$ is the characteristic velocity of eddies on scale $L_c$). Since $\Pran \ll 1$, this scale is also unaffected by viscosity, so one would expect a turbulent cascade with well-defined kinetic energy transfer rate of order $U_c^3 / L_c$. If, in addition, dissipative irreversible conversions into the potential energy reservoir are negligible, then $U_c^3 / L_c = \varepsilon$ where $\varepsilon$ is the viscous energy dissipation rate. Solving for $L_c$ and $U_c$ yields \citep[see][for an alternative derivation of these scalings]{Lignieres2018}: 
\begin{equation} \label{eqn:ZahnLengthScale}
L_c = \left( \frac{\kappa \varepsilon^{1/3} }{N_c^2} \right)^{3/8} \quad \mbox{  and  } \quad U_c^3 = L_c \varepsilon  ,
\label{eq:horizscalings}
\end{equation} 
from which a turbulent diffusion coefficient can then be constructed as
\begin{equation}
D_{\rm turb} \sim U_c L_c \sim \sqrt{\frac{\kappa \varepsilon}{N_c^2} } .
\label{eq:dturbhoriz}
\end{equation}
The \cite{Zahn92} model for turbulent mixing by horizontal shear instabilities at low P\'eclet number and/or low Prandtl number has, to our knowledge, never been tested. In addition to verifying (\ref{eq:horizscalings}) and (\ref{eq:dturbhoriz}), we are also interested in testing the assumption that all energy dissipation is exclusively viscous. Although this assumption is superficially plausible, there is growing evidence \citep{Maffioli2016,Garanaik2019} for flows with $Pr \gtrsim O(1)$ that non-trivial irreversible mixing converting kinetic energy into potential  energy  continues to occur even in the limit $Fr \rightarrow 0$ of extremely strong stratification. 

In what follows we therefore study the simplest possible model of a stratified horizontal shear flow, and focus in this paper on the limit where thermal diffusion is important, or equivalently, the low P\'eclet number limit. This limit is interesting for three reasons.
First, as discussed by \cite{GaraudKulen16}, the thermal diffusivity in the outer layers of the most massive stars (10 $M_{\odot}$ and above) is so large (with $\kappa \sim 10^{11}$m$^2$/s or larger) that the P\'eclet number based on typical stellar length scales and expected flow velocities is smaller than one. Second, even though the global-scale P\'eclet number is large in lower-mass stars or in the deep interiors of high-mass stars (where the thermal diffusivity is much smaller), there must necessarily exist a length scale $L_c$ below which the flow behaves diffusively \citep{Zahn92}, and for which the limit is relevant. Hence, understanding the behaviour of low P\'eclet number flows may provide a way of creating a model for mixing at small scales in stars. Finally, and from a practical perspective, studying high $Pe$ flows with $Pr \ll 1$ is numerically very challenging since it requires very large values of $Re$. As such, understanding the low $Pe$ limit should be viewed as a first step towards the more ambitious goal of understanding low $Pr$ stratified mixing.

Section \ref{sec:model} presents the model setup, and section \ref{sec:linearstability} summarizes the results of a linear stability analysis of the problem. Section \ref{sec:DNS} describes the results of a few characteristic simulations and identifies four separate regimes, each with its own characteristic properties. These are then systematically reviewed in section \ref{sec:nonlinearregimes}, where we study the dominant balances for each regime and derive pertinent scaling laws that are then compared with the numerical data. We discuss these results and draw our conclusions in section \ref{sec:discuss}.


\section{Mathematical formulation} \label{sec:model}

\subsection{Mathematical model}

\begin{figure}
	\centering
	\vspace{0.25cm}
	\includegraphics[width=\linewidth]{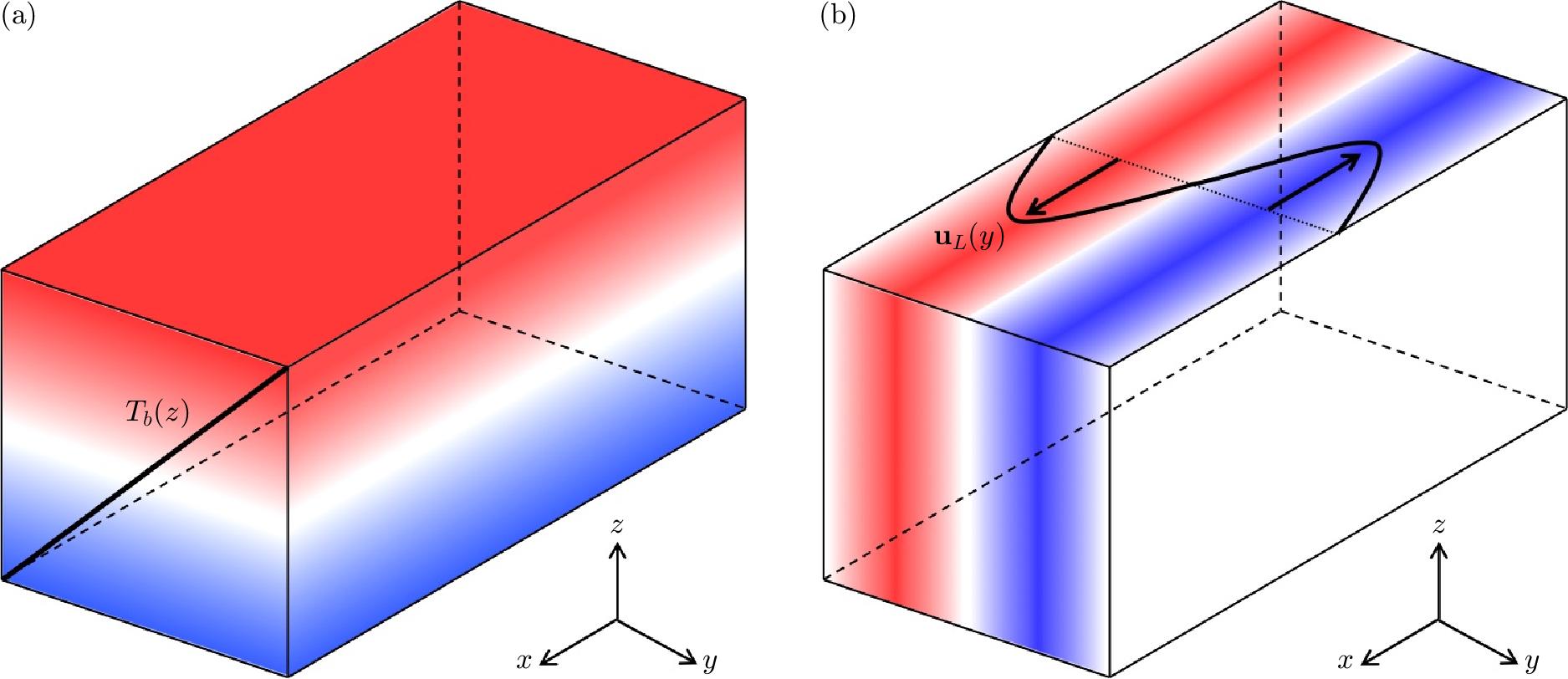}
	\caption{Schematics of the basic state set-up showing (a) the linearized background temperature distribution $T_b(z)$ and (b) the laminar body-forced velocity profile $\mathbf{u}_L(y)$.}
	\label{fig:schematic1}
\end{figure}

We consider an incompressible, body-forced, stably stratified flow with streamwise velocity field aligned with the $x$-axis. In accordance with the Spiegel-Veronis-Boussinesq approximation \citep{SpiegelVeronis1960}, we assume that the basic state comprises a linearized temperature distribution $T_b(z)$ given by $T_b(z)=T_0 + z (dT_b / dz)$, where $T_0$ is a reference temperature, along with a body-forced laminar velocity field $\mathbf{u}_L(y)$. The total temperature field, $T$, includes perturbations $T'(x,y,z,t)$ away from the basic state such that $T = T_b(z) + T'(x,y,z,t)$. As discussed in section \ref{sec:intro2}, the density fluctuations $\rho'$ and temperature fluctuations $T'$ are related according to the linearized equation of state
\begin{equation}
\frac{\rho'}{\rho_0}= - \alpha T',
\end{equation}
where $\rho_0$ is a reference density and $\alpha=- \rho_0^{-1} (\partial \rho / \partial T)$ is the coefficient of thermal expansion. The three-dimensional velocity field is given by $\mathbf{u}(x,y,z,t) = u \mathbf{e}_x + v \mathbf{e}_y + w \mathbf{e}_z$. For numerical efficiency, we impose triply-periodic boundary conditions on the body force $\mathbf{F}$ and the variables $T'$ and $\mathbf{u}$ such that $(x,y,z) \in [0,L_x) \times [0,L_y) \times [0,L_z)$. A suitable candidate for the applied force is a monochromatic sinusoidal forcing driving a horizontal Kolmogorov flow:
\begin{equation}
\mathbf{F} \propto \sin \left ( \frac{2\pi y}{L_y} \right) \mathbf{e}_x.
\end{equation}
This choice of forcing is computationally straightforward to implement and was selected following the work of \cite{Lucas2017} who studied horizontally-sheared stratified flows at $Pr = 1$. The monochromatic Kolmogorov forcing was also used by \cite{Balmforthyoung} to study vertically-sheared stratified flows at high $Pr$, and by \cite{Garaudal15} and \cite{GaraudKulen16} for vertically-sheared stratified flows at low $Pr$ (and in the low $Pe$ limit). It has the advantage of being linearly unstable (as shown below), in contrast with other setups such as the shearing box that only have finite amplitude instabilities. Figure \ref{fig:schematic1} illustrates the basic laminar state.

The governing Spiegel-Veronis-Boussinesq equations \citep{SpiegelVeronis1960} for this model setup are:
\begin{equation}
\frac{\partial \mathbf{u}}{\partial t} + \mathbf{u} \cdot \nabla \mathbf{u} + \frac{1}{\rho_0} \nabla p = \nu \nabla^2 \mathbf{u} + \alpha g T' \mathbf{e}_z + \chi \sin \left ( \frac{2\pi y}{L_y} \right) \mathbf{e}_x,
\end{equation}
\begin{equation}
\frac{\partial T'}{\partial t} + \mathbf{u} \cdot \nabla T' +  w \left ( \frac{d T_b}{dz} + \frac{g}{c_p} \right ) = \kappa \nabla^2 T',
\end{equation}
\begin{equation}
\nabla \cdot \mathbf{u} = 0,
\end{equation}
where $\nu$ is the kinematic viscosity, $\kappa$ is the thermal diffusivity, $\chi$ is the forcing amplitude, $p$ is the pressure, $c_p$ is the specific heat at constant pressure and gravity $g$ acts in the negative $z$-direction. In this study, we specify that $L_y = L_z$ while $L_x$ may vary continuously such that the aspect ratio of the domain is given by $\lambda = L_x / L_y$. The case $\lambda > 1$ corresponds to domains which are longer in the streamwise direction.

\subsection{Non-dimensionalization and model parameters}

In equilibrium, we anticipate a balance between the body force and fluid inertia such that $\mathbf{u} \cdot \nabla \mathbf{u} \sim \chi \sin \left ( 2\pi y/L_y \right) \mathbf{e}_x$ in the streamwise direction. For a characteristic length scale $L_y/2\pi$, this gives a characteristic velocity scale $\sqrt{ \chi L_y / 2\pi}$ and a characteristic time scale $\sqrt{L_y/2\pi \chi}$. Combined with the vertical temperature gradient scale $(dT_b / dz + g/c_p)$, we use the equivalent non-dimensionalization as in \cite{Lucas2017} to give the following system of equations, in which all quantities are non-dimensional:
\begin{equation} \label{eqn:momentumnondim}
\frac{\partial \mathbf{u}}{\partial t} + \mathbf{u} \cdot \nabla \mathbf{u} +  \nabla p = \frac{1}{Re} \nabla^2 \mathbf{u} + B T' \mathbf{e}_z + \sin (y) \mathbf{e}_x,
\end{equation}
\begin{equation} \label{eqn:densitynondim}
\frac{\partial T'}{\partial t} + \mathbf{u} \cdot \nabla T' + w = \frac{1}{RePr} \nabla^2 T',
\end{equation}
\begin{equation}  \label{eqn:continuitynondim}
\nabla \cdot \mathbf{u} = 0.
\end{equation}
We thus have three non-dimensional numbers: the Reynolds number $Re$; the buoyancy parameter $B$; and the Prandtl number $Pr$, which determine the dynamics of the system:
\begin{equation}
Re := \frac{\sqrt{\chi}}{\nu}\left( \frac{L_y}{2\pi} \right)^{\frac{3}{2}}, \quad B := \frac{\alpha g (dT_b / dz + g / c_p) L_y}{2\pi \chi} = \frac{N_b^2 L_y}{2\pi \chi}, \quad Pr := \frac{\nu}{\kappa},
\label{eq:BRedef}
\end{equation}
where $N_b$ is the dimensional buoyancy frequency defined in (\ref{eq:ncstardef}), which is now constant by construction. Note that $B$ is related to the Froude number as
\begin{equation}
B = Fr^{-2}.
\end{equation}
It is also convenient to introduce the P\'{e}clet number $Pe$, defined as 
\begin{equation}
Pe := Re Pr = \frac{\sqrt{\chi}}{\kappa}\left( \frac{L_y}{2\pi} \right)^{\frac{3}{2}}.
\label{eq:Pedef}
\end{equation}
Both sets of parameters, ($Re$, $B$, $Pr$) or ($Re$, $B$, $Pe$), uniquely define the system and will be used interchangeably throughout this study. In all that follows, the domain is a cuboid such that $(x,y,z) \in [0,{2\pi\lambda}) \times [0,{2\pi}) \times [0,{2\pi})$, and variables $p$, $T'$ and $\mathbf{u}$ have triply-periodic boundary conditions. This system, defined by (\ref{eqn:momentumnondim}), (\ref{eqn:densitynondim}) and (\ref{eqn:continuitynondim}), will henceforth be referred to as the standard system of equations.

\subsection{Low-P\'{e}clet number approximation} \label{subsec:lpnmodel}

As discussed in section \ref{sec:intro2}, when the thermal diffusion timescale is much shorter than the advective timescale, a \emph{quasi-static} regime is established where temperature fluctuations are slaved to the vertical velocity field. Motivated by the astrophysical applications described in section \ref{sec:intro2}, we consider the standard set of equations (\ref{eqn:momentumnondim})-(\ref{eqn:continuitynondim}) in the asymptotic limit of low P\'{e}clet number (LPN). This limit was studied by \citet{Spiegel1962} and \cite{Thual1992} in the context of thermal convection, and more recently by \citet{Lignieres1999} in the context of stably stratified flows. Ligni\`{e}res proposed that the standard equations can be approximated by a reduced set of equations called the ``low-P\'{e}clet number" equations (LPN equations hereafter), in which the density fluctuations are slaved to the vertical velocity field:
\begin{equation}
\frac{\partial \mathbf{u}}{\partial t} + \mathbf{u} \cdot \nabla \mathbf{u} +  \nabla p = \frac{1}{Re} \nabla^2 \mathbf{u} + B T' \mathbf{e}_z + \sin (y)  \mathbf{e}_x,
\end{equation}
\begin{equation}\label{eqn:LPNdensity}
w - \frac{1}{Pe} \nabla^2 T' = 0,
\end{equation}
\begin{equation}
\nabla \cdot \mathbf{u} = 0.
\end{equation}
These can be derived by assuming a regular asymptotic expansion of $T'$ in powers of $Pe$, i.e. $T'=T'_0 + T'_1 Pe + O(Pe^2)$, and by assuming that the velocity field is of order unity. At lowest order ($Pe^{-1}$), we get $\nabla^2 T'_0=0$ implying that $T'_0=0$ is required to satisfy the boundary conditions, while at the next order ($Pe^{0}$), the equations yield $w = \nabla^2 T'_1 \approx Pe^{-1}\nabla^2 T'$ as required.

Noting that (\ref{eqn:LPNdensity}) can be re-written formally as $T'=Pe \nabla^{-2}w$, we derive the reduced set of LPN equations:
\begin{equation}\label{eqn:lpnequation1}
\frac{\partial \mathbf{u}}{\partial t} + \mathbf{u} \cdot \nabla \mathbf{u} +  \nabla p = \frac{1}{Re} \nabla^2 \mathbf{u} + BPe \nabla^{-2}w \mathbf{e}_z + \sin (y)  \mathbf{e}_x,
\end{equation}
\begin{equation}\label{eqn:lpnequation2}
\nabla \cdot \mathbf{u} = 0.
\end{equation}
These equations explicitly demonstrate that under the LPN approximation (and in contrast to the standard equations), there are only two non-dimensional parameters governing the flow dynamics, notably the Reynolds number $Re$ and the product of the buoyancy parameter and the P\'{e}clet number, $BPe = Pe Fr^{-2}$. This combined parameter, which we consider to be a measure of the stratification, can take any value (even for small P\'{e}clet numbers) because $B$ can be arbitrarily large, or equivalently $Fr$ can be arbitrarily small, as the stratification becomes strong. 

There are advantages of studying the LPN equations rather than the standard equations. For example, this reduced set of equations allows for the derivation of mathematical results such as an energy stability threshold that explicitly depends on $BPe$ \citep[see][]{Garaudal15}. 
Throughout this study, we will discuss both systems of equations, verifying the validity of the LPN equations where possible. 


\section{Linear stability analysis} 
\label{sec:linearstability}

\subsection{Standard equations}

We begin by considering the stability of a laminar flow to infinitesimal perturbations, with initial focus on the standard set of equations (\ref{eqn:momentumnondim})-(\ref{eqn:continuitynondim}). The background flow $\mathbf{u}_L(y)$, which satisfies $Re^{-1}\nabla^2 \mathbf{u}_L + \sin (y) \mathbf{e}_x =0$, is given by
\begin{equation} \label{eqn:laminarsolution}
\mathbf{u}_L(y) = Re \sin (y) \mathbf{e}_x.
\end{equation}
Note that if one wishes to consider a basic state with generic amplitude $aRe$ instead of amplitude $Re$, it is straightforward to apply a rescaling using the method described in appendix \ref{appA}.
For small perturbations $\mathbf{u}'(x,y,z,t)$ away from this laminar flow, i.e. letting $\mathbf{u} = \mathbf{u}_L(y) + \mathbf{u}'(x,y,z,t)$, the linearised perturbation equations are:
\begin{equation}\label{eqn:linear1}
\frac{\partial \mathbf{u}'}{\partial t} + Re   \cos (y) v' \mathbf{e}_x + Re \sin(y) \frac{\partial \mathbf{u}'}{\partial x} + \nabla p = \frac{1}{Re} \nabla^2 \mathbf{u}' + BT' \mathbf{e}_z,
\end{equation}
\begin{equation}\label{eqn:linear2}
\frac{\partial T'}{\partial t} + Re \sin(y) \frac{\partial T'}{\partial x} +  w' = \frac{1}{RePr} \nabla^2 T',
\end{equation}
\begin{equation}\label{eqn:linear3}
\nabla \cdot \mathbf{u}' = 0.
\end{equation}
In this set of partial differential equations (PDEs), the coefficients are periodic in $y$ but independent of $x$, $z$ and $t$. Consequently, and in the conventional fashion, we consider normal mode disturbances of the form: 
\begin{equation}
q(x,y,z,t) = \hat{q}(y) \exp[ik_x x + ik_z z + \sigma t],
\end{equation}
where $q \in (u', v', w', T', p)$ and $k_x$ and $k_z$ are the perturbation wavenumbers in the $x$ and $z$-directions respectively. The geometry of the model set-up requires that $k_x \in \mathbb{R}$ and $k_z \in \mathbb{Z}$. We seek periodic solutions for $\hat{q}(y)$ given by
\begin{equation}
\hat{q}(y) = \sum_{l=-L}^{L} q_l e^{ily}.
\end{equation}
Substituting this ansatz into equations (\ref{eqn:linear1})-(\ref{eqn:linear3}) and using the orthogonality property of complex exponentials, we obtain a $5 \times (2L+1) = (10L+5)$ algebraic system of equations for the $u_l$, $v_l$, $w_l$, $T_l$ and $p_l$ for $l \in (-L,L)$:
\begin{equation} \label{eqn:lsstandard1}
\frac{1}{2}Re k_x (u_{l+1} -u_{l-1}) - \frac{l^2 + k_x^2 + k_z^2}{Re} u_l - \frac{1}{2}Re(v_{l-1} + v_{l+1}) - ik_x p_l = \sigma u_l,
\end{equation}
\begin{equation} \label{eqn:lsstandard2}
\frac{1}{2}Re k_x (v_{l+1} -v_{l-1}) - \frac{l^2 + k_x^2 + k_z^2}{Re} v_l - il p_l = \sigma v_l,
\end{equation}
\begin{equation} \label{eqn:lsstandard3}
\frac{1}{2}Re k_x (w_{l+1} -w_{l-1})- \frac{l^2 + k_x^2 + k_z^2}{Re} w_l + B T_l - ik_z p_l = \sigma w_l,
\end{equation}
\begin{equation} \label{eqn:lsstandard4}
\frac{1}{2}Re k_x (T_{l+1} -T_{l-1}) - w_l - \frac{l^2 + k_x^2 + k_z^2}{RePr} T_l = \sigma T_l,
\end{equation}
\begin{equation} \label{eqn:lsstandard5}
k_x u_l + lv_l + k_z w_l = 0.
\end{equation}
This system can be re-formulated as a generalised eigenvalue problem for the complex growth rates $\sigma$,
\begin{equation} \label{eqn:linearstab}
\mathbf{A}(k_x,k_z,Re,B,Pr) \mathbf{X} = \sigma \mathbf{B}\mathbf{X},
\end{equation}
where $\mathbf{X} = (u_{-L},...,u_L,v_{-L},...,v_L,w_{-L},...,w_L,T_{-L},...,T_L,p_{-L},...,p_L)$, $\mathbf{A}$ and $\mathbf{B}$ are $(10L+5)\times (10L+5)$ square matrices and $\mathbf{B}_{i,j}= \{\delta_{ij}, \  i,j \leq (8L+4); \ 0, \ \textrm{otherwise}\}$. Equation (\ref{eqn:linearstab}) has $(10L+5)$ eigenvalues $\sigma$. For perturbation wavenumbers $k_x$ and $k_z$ and system parameters $Re$, $B$ and $Pr$, the eigenvalue with the largest real part determines the growth rate of the linear instability. The eigenvalue problem can be solved numerically, with $L$ chosen such that convergence is achieved. 

\subsubsection{Comparison with previous results at $Pr=1$}


We first consider the case of $Pr=1$ for ease of comparison with previous work. \citet{Deloncle2007}, \citet{AroboneSarkar2012} and \citet{Parkal2020} each considered the linear stability of horizontal shear layers with somewhat different base flows, and \citet{Lucas2017} considered the linear stability of the specific horizontally-sheared Kolmogorov flow considered here, exclusively for $Pr=1$. Letting $B=100$, we consider the linear stability of the basic state flow $\mathbf{u}_L$ (see (\ref{eqn:laminarsolution})) across a range of Reynolds numbers for both 2D and 3D perturbation modes. 

\begin{figure}
	\centering
	\vspace{0.25cm}
	\includegraphics[width=\linewidth]{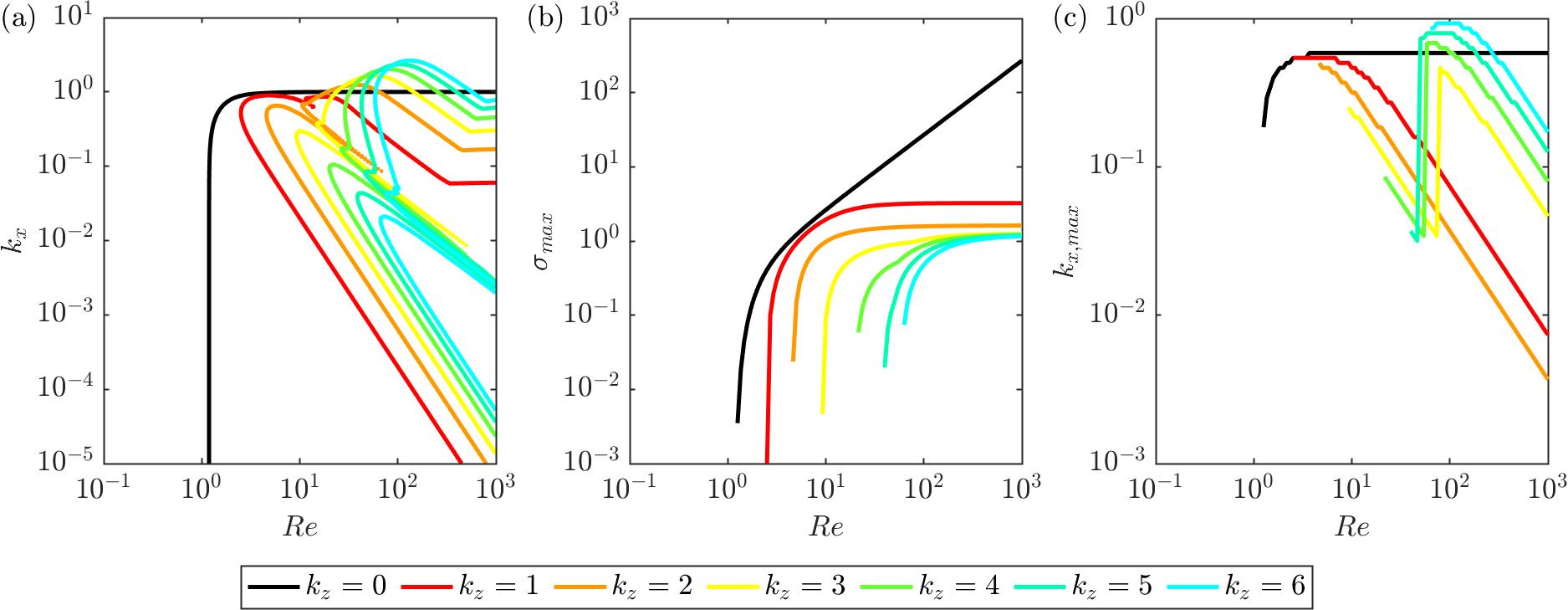}
	\caption{(a) Neutral stability curves for a range of $k_z$ wavenumbers as a function of Reynolds number and $k_x$ wavenumber, with instability occurring to the right and below the curves. Variation with Reynolds number for a collection of $k_z$ wavenumbers of: (b) the largest growth rate $\sigma_{max}$ maximised across all horizontal wavenumbers $k_x$; (c) the associated horizontal wavenumber $k_{x,max}$. The curves plotted include $k_z=0$ (black) and $k_z =1, 2, 3, 4, 5, 6$ (coloured) and the standard equations were used with $B=100$ and $Pr=1$ fixed (so $Pe=Re$).}
	\label{fig:linearkzsigma}
\end{figure}

Figure \ref{fig:linearkzsigma}(a) shows the neutral stability curves ($\sigma=0$) for varying vertical wavenumbers $k_z \in (0,...,6)$ in the $(Re,k_x)$ space. Our results are in agreement with those of \citet{Lucas2017}. Stability ($\sigma < 0$) is found to the left and above the curves whilst instability ($\sigma > 0$) occurs to the right and below. The black curve illustrates the 2D ($k_z=0$) mode. This neutral stability curve intercepts the $x$-axis when $Re=2^{1/4}\simeq 1.19 $, implying that the system is linearly stable when $Re<2^{1/4}$ \citep[in agreement with][once the correct rescaling is applied (see appendix \ref{appB} for details)]{Beaumont1981,Balmforthyoung}. For large $Re$, it asymptotes to $k_x=1$ but, in agreement with \cite{Lucas2017}, always lies below this line, leading to the conclusion that domains such that $\lambda = L_x / L_y \leq 1$ are linearly stable to the 2D mode.

The coloured curves show the neutral stability curves for the first six 3D modes ($k_z \in (1,...,6)$). The onset of instability in the 3D modes is found to occur for higher Reynolds numbers than the 2D mode, with the critical Reynolds number for instability of these 3D modes increasing monotonically with increasing $k_z$. For a range of $Re \sim O(100)$ (corresponding to $Pe \sim O(100)$), the 3D curves actually cross the line $k_x=1$ implying that these modes are unstable for domains where $\lambda =1$, i.e. cubic domains. 

Figures \ref{fig:linearkzsigma}(b)-(c) further analyse the information in figure \ref{fig:linearkzsigma}(a) by computing, for each Reynolds number and $k_z$, the largest (positive) growth rate, $\sigma_{max}$, across all values of $k_x$ and the value of $k_x$ for which that maximum is achieved, $k_{x,max}$. We see that, as well as being the mode that becomes unstable first, the 2D mode is always the fastest growing one. In addition, the ratio of the growth rate of the 2D mode to that of the 3D modes increases with $Re$. We therefore predict that the 2D mode would strongly influence the dynamics when it is unstable (i.e. for domain sizes such that $\lambda>1$). Finally, we note that the corresponding streamwise wavenumbers of the fastest growing 3D modes satisfy $k_{x,max} \to 0$ in the limit $Re \to \infty$, while those of the fastest growing 2D mode remain constant.


\subsubsection{Stability at low $Pr$}

\begin{figure}
	\centering
	\vspace{0.25cm}
	\includegraphics[width=\linewidth]{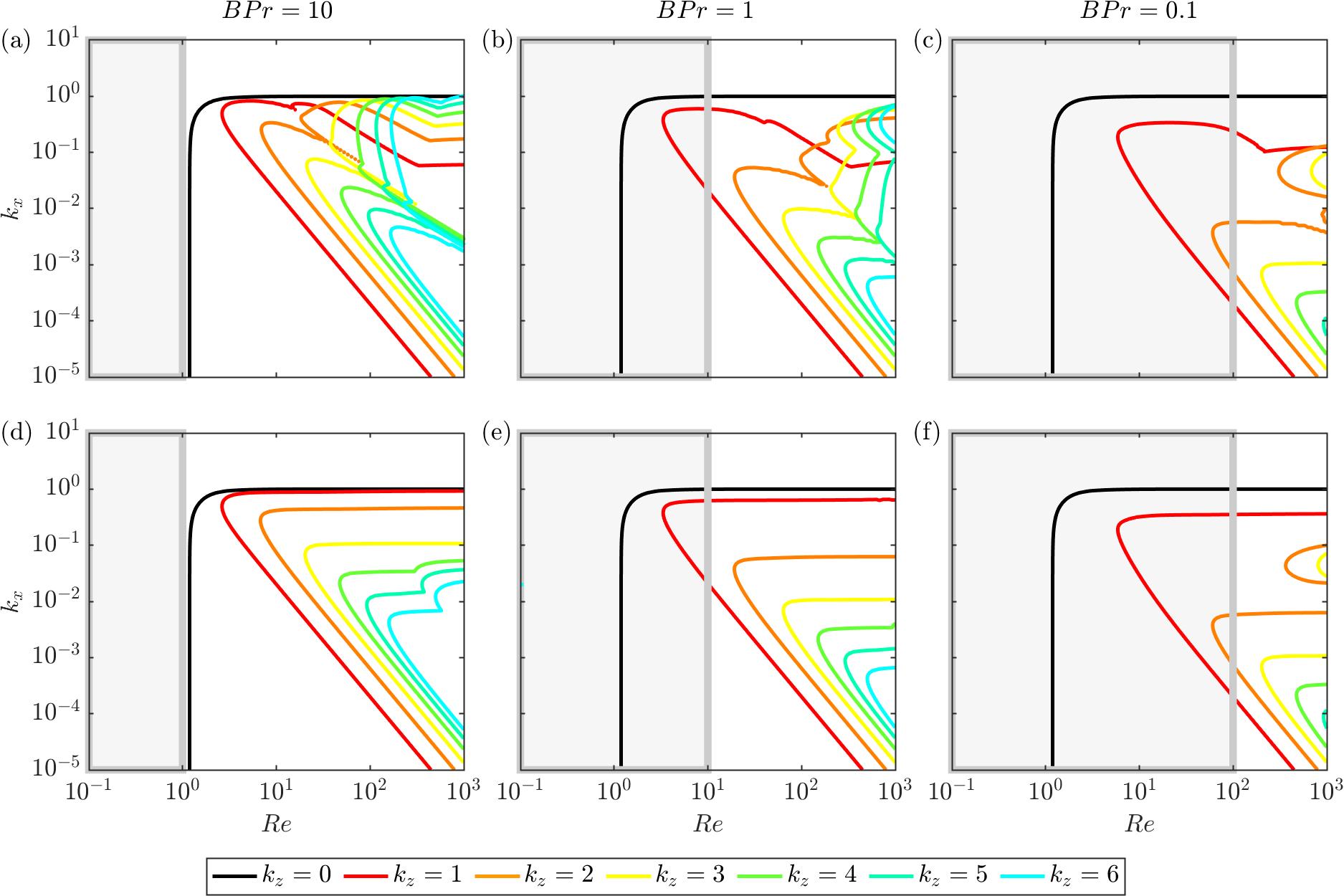}
	\caption{A comparison of linear stability analysis results between the standard equations (top row) and the LPN equations (bottom row). Neutral stability curves for a range of $k_z$ wavenumbers ($k_z=0$ (black) and $k_z =1, 2, 3, 4, 5, 6$ (coloured)) are plotted as a function of Reynolds number and $k_x$. Instability occurs to the right and below the curves. Parameter values used are (a) $B=100$, $Pr=0.1$, (b) $B=100$, $Pr=0.01$, (c) $B=100$, $Pr=0.001$, (d) $BPr=10$, (e) $BPr=1$, (f) $BPr=0.1$. Grey rectangles indicate regions where $Pe \le 0.1$.} 
	\label{fig:linearfulllpn}
\end{figure}

Astrophysical applications motivate an understanding of the effects of the stratification parameter $B$ and Prandtl number $Pr$ on the linear stability of the basic state. Consequently, in figures \ref{fig:linearfulllpn}(a)-(c) we plot the neutral stability curves in exactly the same fashion as in figure \ref{fig:linearkzsigma}(a), for three different Prandtl numbers: $Pr=0.1$ (first column), $Pr=0.01$ (second column) and $Pr=0.001$ (third column), keeping $B=100$ constant. Whilst the neutral stability curves for the 2D mode are identical, clear trends exist for the 3D modes. A reduction in the value of $Pr$ shifts the critical Reynolds numbers for the onset of instability of the 3D modes towards higher values, thereby making these modes less unstable. This result is consistent with \citet{AroboneSarkar2012} and \citet{Parkal2020}, who investigated the stability of a diffusive, stratified, horizontally-sheared hyperbolic flow. We also note that the same trend is found by letting $B \to 0$ and keeping $Pr$ constant (not plotted). Thus, $B \to 0$ (at fixed $Pr$) and $Pr \to 0$ (at fixed $B$) have the same effect: the 3D modes of instability are suppressed while the 2D mode remains unstable. The explanation for this emerges from consideration of (\ref{eqn:densitynondim}). As the Prandtl number tends to zero (keeping the Reynolds number finite), the P\'{e}clet number becomes small and so the buoyancy diffusion becomes important. In this case, a parcel of fluid that is advected into surrounding fluid of a different density adjusts very rapidly to its surroundings, thereby reducing the buoyancy force and so approximating an unstratified system.

However, it is important to note that another distinguished limit exists in which $B \to \infty$ and $Pr \to 0$, while the product $BPr$ remains finite. This limit is relevant to stellar interiors, and behaves quite differently from the case where $B$ is fixed while $Pr \to 0$, as we now demonstrate.


\subsection{Low-P\'{e}clet number equations}

We now examine the linear stability of the LPN equations, given by (\ref{eqn:lpnequation1}) and (\ref{eqn:lpnequation2}). We follow the same steps as in the previous section, however we find ourselves working this time with a reduced set of four equations rather than five. We obtain a $4 \times (2L+1) = (8L+4)$ algebraic system of equations for the $u_l$, $v_l$, $w_l$ and $p_l$ for $l \in (-L,L)$:
\begin{equation} \label{eqn:lslpn1}
\frac{1}{2}Re k_x (u_{l+1} -u_{l-1}) - \frac{l^2 + k_x^2 + k_z^2}{Re} u_l - \frac{1}{2}Re(v_{l-1} + v_{l+1}) - ik_x p_l = \sigma u_l,
\end{equation}
\begin{equation} \label{eqn:lslpn2}
\frac{1}{2}Re k_x (v_{l+1} -v_{l-1}) - \frac{l^2 + k_x^2 + k_z^2}{Re} v_l - il p_l = \sigma v_l,
\end{equation}
\begin{equation} \label{eqn:lslpn3}
\frac{1}{2}Re k_x (w_{l+1} -w_{l-1})- \frac{l^2 + k_x^2 + k_z^2}{Re} w_l - \frac{BPe}{k_x^2 + k_z^2} w_l - ik_z p_l = \sigma w_l,
\end{equation}
\begin{equation} \label{eqn:lslpn4}
k_x u_l + lv_l + k_z w_l = 0.
\end{equation}
As before, this can be re-formulated as a generalised eigenvalue problem for the complex growth rates $\sigma$,
\begin{equation}
\mathbf{A}(k_x,k_z,Re,BPe) \mathbf{X} = \sigma \mathbf{B}\mathbf{X},
\end{equation}
where $\mathbf{X} = (u_{-L},...,u_L,v_{-L},...,v_L,w_{-L},...,w_L,p_{-L},...,p_L)$, $\mathbf{A}$ and $\mathbf{B}$ are $(8L+4)\times (8L+4)$ square matrices and $\mathbf{B}_{i,j}= \{\delta_{ij}, \  i,j \leq (6L+3); \ 0, \ \textrm{otherwise}\}$. We follow the same procedure as before, solving the eigenvalue problem numerically. 


In order to test the validity of the LPN equations, we first compare the results of the linear stability analysis in the LPN limit to that obtained using the standard equations. Figure \ref{fig:linearfulllpn} illustrates this comparison. The top row (as already discussed) shows the neutral stability curves from the standard equations and the bottom row shows the equivalent results from the LPN equations. The value of $BPr = BPe/Re$ in the bottom row decreases from left to right by two orders of magnitude, in line with reductions in the value of $BPe$ at fixed $Re$ in the standard equations in the top row. 
As demonstrated in section \ref{subsec:lpnmodel}, the LPN equations are asymptotically correct in the limit where $Pe \to 0$. Figure \ref{fig:linearfulllpn} shows that they remain valid up to $Pe \simeq 0.1$ (i.e. within the regions shown in grey). Outside of these regions, increasingly large differences emerge, especially as $Pe$ increases above one. 
In particular, the neutral stability curves for the 3D modes never cross the line $k_x=1$ in the LPN equations, suggesting that horizontal shear instabilities 
do not arise for cubic domains when the LPN approximation is used. 

\begin{figure}
	\centering
	\vspace{0.25cm}
	\includegraphics[width=\linewidth]{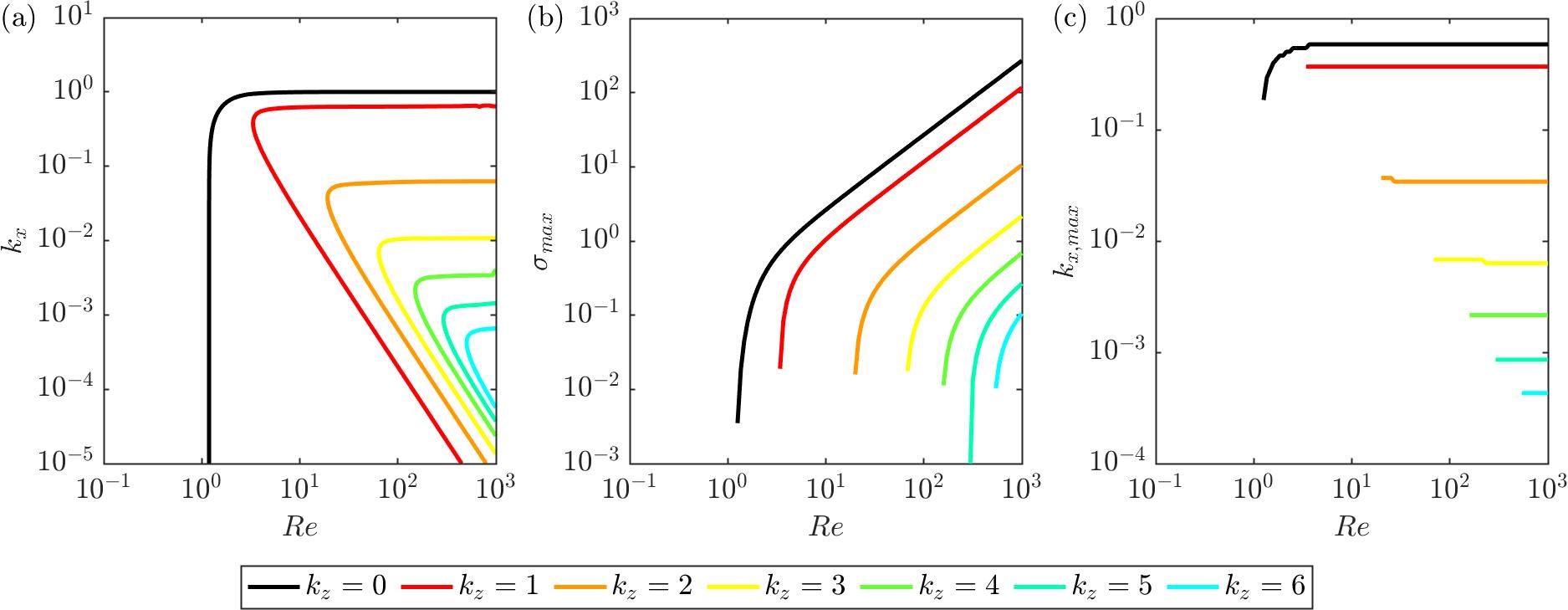}
	\caption{(a) Neutral stability curves for a range of $k_z$ wavenumbers as a function of $Re$ and $k_x$, with instability occurring to the right and below the curves. This time we used the LPN equations, with $BPr=1$ fixed. Variation with Reynolds number for a collection of $k_z$ wavenumbers of: (b) the largest growth rate $\sigma_{max}$ maximised across all horizontal wavenumbers $k_x$; (c) the associated horizontal wavenumber $k_{x,max}$. The curves plotted include $k_z=0$ (black) and $k_z =1, 2, 3, 4, 5, 6$ (coloured).}
	\label{fig:linearkzsigmalpn}
\end{figure}


The LPN system of equations depend on the combined parameter $BPr = BPe/Re$. In the limit of strong stratification ($B \to \infty$) and strong thermal diffusion ($Pr \to 0$), this parameter remains finite and is not necessarily small. As can be seen in figure \ref{fig:linearfulllpn}, the 3D modes remain unstable in this limit, in agreement with the results from the standard system of equations.

We now focus on the case when $BPr=1$. By way of comparison with the standard equations at $Pr=1$, figures \ref{fig:linearkzsigmalpn}(b)-(c) show, for each Reynolds number and $k_z$ wavenumber, the largest (positive) growth rate, $\sigma_{max}$, across all values of $k_x$ and the value of $k_x$ for which that maximum is achieved, $k_{x,max}$. As before, we observe that the 2D mode is both the first mode to become unstable, and is always the fastest growing mode. There are, however, two significant differences between high and low Prandtl number dynamics. Firstly, in the LPN limit, figure \ref{fig:linearkzsigmalpn}(b) shows that the growth rates of the fastest growing 3D modes increase in line with those of the fastest growing 2D mode. Secondly, the corresponding values of $k_{x,max}$ remain constant as $Re \to \infty$. Consequently, the 3D modes remain important relative to the 2D mode and we therefore predict that, in contrast to the case when $Pr=1$, both the 2D and 3D modes would strongly influence the dynamics in this limit. These results, combined with the fact that the 3D modes remain unstable in the limit of strong stratification and strong thermal diffusion, have important consequences, as we shall see in section \ref{sec:DNS}. 



\section{Direct numerical simulations}
\label{sec:DNS}

We now present results from a series of DNSs of horizontal shear flows at low P\'eclet number following the model setup and equations described in section \ref{sec:model}. As we shall demonstrate, the system presents a rich ecosystem of instabilities that feed on each other, leading to a number of distinct dynamical regimes that will be further characterized in section \ref{sec:nonlinearregimes}.

\subsection{Numerical algorithm}

The DNSs are performed using the PADDI code first introduced by \citet{Traxleretal2011b} and \citet{Stellmachetal2011} to study double-diffusive fingering. The code has since then been modified to study many different kinds of instabilities, including body-forced vertical shear instabilities, using both the standard equations and the LPN approximation \citep{Garaudal15,GaraudKulen16,GagnierGaraud2018,KulenGaraud2018}. PADDI is a triply-periodic pseudo-spectral algorithm that uses pencil-based Fast Fourier Transforms, and third order Backward-Differentiation Adams-Bashforth adaptive timestepping \citep{peyret2002smi} in which diffusive terms are treated implicitly while all other terms are treated explicitly. The velocity field is made divergence-free at every timestep by solving the relevant Poisson equation for the pressure. Two versions of the code exist, one that solves the standard equations (\ref{eqn:momentumnondim})-(\ref{eqn:continuitynondim}), and one that solves the LPN equations (\ref{eqn:lpnequation1})-(\ref{eqn:lpnequation2}). 

Based on the linear stability analysis performed in section \ref{sec:linearstability}, we have selected a domain size such that $L_y = L_z = 2\pi$ and $L_x = 4\pi$. This allows for the natural development of a single 2D mode of instability (for which $k_x = 0.5$),  without being computationally prohibitive at high Reynolds number (see below). 
A comparison of simulation outcomes for different domain lengths is presented in \citet[section 4.2]{CopeGFD} but only for $Re = 50$, for which only two dynamical regimes exist. A systematic exploration of the effect of domain aspect ratio at high Reynolds number will be the subject of future work. 
\begin{table}
\begin{center}
\begin{tabular}{cccccccc}
$Re$ & $Pe$ & $B$ & $Re_{\lambda}$ & $l_z \pm \delta l_z$ & $w_{rms}\pm \delta w_{rms}$  & $T^{\prime}_{rms}\pm \delta T^{\prime}_{rms}$  & $\eta \pm \delta \eta$  \vspace{0.15cm} \\

$600$	&$0.1$	&$6000$	&$280$	&$0.25\pm 0.02$	&$0.24\pm 0.03$	& $(3.5 \pm 0.3)\times 10^{-4}$	&$0.36\pm 0.03$	\\
$600$	&$0.1$	&$12\,000$	&$359$	&$0.19\pm 0.02$	&$0.21\pm 0.03$	&$(2.3\pm 0.3) \times 10^{-4}$	&$0.31\pm 0.02$	 \\
$600$	&$0.1$	&$100\,000$ &$465$	&$0.11\pm 0.005$	&$0.038\pm 0.003$	&$(4.0\pm 0.4) \times 10^{-5}$ &$0.18 \pm 0.02$\vspace{0.15cm} \\ 

$300$	&$0.1$	&$1$	&$173$	&$1.88\pm 0.29$	&$0.91\pm 0.10$		&$(3.1\pm 0.8)\times 10^{-2}$	&$0.029\pm 0.007$ \\
$300$	&$0.1$	&$100$	&$171$	&$0.99\pm 0.08$	&$0.56\pm 0.03$		&$(6.9\pm 0.9)\times 10^{-3}$	&$0.43\pm 0.05$	\\
$300$	&$0.1$	&$1000$	&$173$	&$0.45\pm 0.02$	&$0.34\pm 0.03$		&$(1.4\pm 0.1)\times 10^{-3}$	&$0.42\pm 0.03$	\\
$300$	&$0.1$	&$3000$	&$227$	&$0.32\pm 0.05$	&$0.25\pm 0.03$		&$(6.5\pm 0.6)\times 10^{-4}$	&$0.35\pm 0.02$	\\
$300$	&$0.1$	&$6000$	&$300$	&$0.24\pm 0.01$	&$0.17\pm 0.03$		&$(4.1\pm 0.6)\times 10^{-4}$	&$0.30\pm 0.03$	\\
$300$	&$0.1$	&$10\,000$ &$319$	&$0.19\pm 0.02$	&$0.11\pm 0.04$		&$(2.7\pm 0.7)\times 10^{-4}$	&$0.23\pm 0.04$	\\
$300$	&$0.1$	&$30\,000$	&$281$ &$0.15\pm 0.01$	&$0.05\pm 0.004$ &$(9.7\pm 0.8)\times 10^{-5}$	&$0.18\pm 0.02$	\\
$300$	&$0.1$	&$100\,000$	&$265$  &$0.13\pm 0.01$	&$0.03\pm 0.003$  &$(4.2\pm	0.4)\times 10^{-5}$	&$0.16\pm 0.02$	\\
$300$	&$0.1$	&$300\,000$	&$219$  &$0.11\pm 0.000$ &$0.02\pm 0.002$ &$(2.1\pm 0.2)\times 10^{-5}$	&$0.15\pm 0.01$	\vspace{0.15cm} \\

$100$	&$0.01$	&$1$	&$102$	&$1.94\pm0.27$	&$0.91\pm 0.11$		&$(5.3\pm 2.2)\times 10^{-3}$	&$0.004\pm 0.002$	\\
$100$	&$0.01$	&$10$	&$98$	&$1.94\pm0.34$	&$0.89\pm 0.09$		&$(3.3\pm 0.9)\times 10^{-3}$	&$0.031\pm 0.007$	\\
$100$	&$0.01$	&$100$	 &$91$	&$1.62\pm0.14$	&$0.83\pm 0.09$		&$(2.0\pm 0.3)\times 10^{-3}$	&$0.18\pm 0.04$	\\
$100$	&$0.1$	&$100$		&$93$	&$0.96\pm0.07$	&$0.49\pm 0.06$		&$(7.5\pm 1.0)\times 10^{-3}$	&$0.45\pm 0.04$	\\
$100$	&$0.1$	&$1000$	&$201$	&$0.39\pm0.03$	&$0.17\pm 0.03$		&$(1.4\pm 0.2)\times 10^{-3}$	&$0.26\pm 0.04$	\\
$100$	&$0.1$	&$3000$	&$161$	&$0.28\pm0.02$	&$0.09\pm 0.01$		&$(5.4\pm 0.8)\times 10^{-4}$	&$0.20\pm 0.03$	\\
$100$	&$0.1$	&$10\,000$	&$172$	&$0.22\pm0.02$	&$0.06\pm 0.01$		&$(2.2\pm 0.5)\times 10^{-4}$	&$0.15\pm 0.03$	\\
$100$	&$0.1$	&$100\,000$	&$112$  &$0.16\pm0.004$ &$0.02\pm 0.001$ &$(4.4\pm 0.3)\times 10^{-5}$	&$0.105\pm 0.006$ \\
$100$	&$1$	&$100$		&$152$	&$0.42\pm 0.03$	&$0.20\pm 0.035$		&$(1.6\pm 0.3)\times 10^{-2}$	&$0.29\pm 0.04$	\\
$100$	&$1$	&$300$		&$165$	&$0.28\pm 0.02$	&$0.08\pm 0.011$		&$(5.2\pm 0.8)\times 10^{-3}$	&$0.19\pm 0.03$	\\
$100$	&$1$	&$500$		&$151$	&$0.26\pm 0.01$	&$0.06\pm 0.006$		&$(3.4\pm 0.4)\times 10^{-3}$	&$0.17\pm 0.02$	\\
$100$	&$1$	&$1000$	&$140$	&$0.23\pm 0.000$	&$0.05\pm 0.005$		&$(2.2\pm 0.3)\times 10^{-3}$	&$0.17\pm 0.02$	\\
$100$	&$1$	&$10\,000$	&$117$	&$0.18\pm 0.02$	&$0.015\pm 0.001$		&$(4.5\pm 0.5)\times 10^{-4}$	&$0.11\pm 0.01$	\\
$100$	&$1$	&$30\,000$	&$245$	&$0.19\pm 0.02$	&$0.007\pm 0.002$		&$(2.7\pm 0.7)\times 10^{-4}$	&$0.09\pm 0.03$	\\
$100$	&$1$	&$50\,000$	&$362$	&$0.20\pm 0.03$	&$0.005\pm 0.002$		&$(1.9\pm 0.6)\times 10^{-4}$	&$0.08\pm 0.04$	\\
$100$	&$1$	&$100\,000$	&$437$	&$0.21\pm 0.02$	&$0.003\pm 0.001$		&$(1.2\pm 0.4)\times 10^{-4}$	&$0.05\pm 0.02$	\\
$100$	&$1$	&$1\,000\,000$ &$554$ &$0.30\pm 0.04$ &$0.0002\pm 0.00004$ &$(1.2\pm 0.2)\times 10^{-5}$ &$0.005\pm 0.002$ \vspace{0.15cm} \\

$50$	&$0.1$	&$0.3$		&$68$	&$2.05\pm 0.30$	&$0.85\pm 0.11$	&$(4.5\pm 1.8)\times 10^{-2}$  &$0.013\pm 0.004$	\\
$50$	&$0.1$	&$1$		&$68$	&$2.02\pm 0.37$	&$0.82\pm 0.09$	&$(3.6\pm 0.9)\times 10^{-2}$	&$0.04\pm 0.008$	\\
$50$	&$0.1$	&$10$		&$63$	&$1.59\pm 0.17$	&$0.71\pm 0.09$		&$(2.0\pm 0.4)\times 10^{-2}$	&$0.19\pm 0.03$	\\
$50$	&$0.1$	&$30$		&$64$	&$1.28\pm 0.13$	&$0.55\pm 0.06$		&$(1.3\pm 0.2)\times 10^{-2}$	&$0.32\pm 0.04$	\\
$50$	&$0.1$	&$100$		&$67$	&$0.89\pm 0.07$	&$0.39\pm 0.05$		&$(7.2 \pm 1.2)\times 10^{-3}$	&$0.33\pm 0.04$	\\
$50$	&$0.1$	&$300$		&$102$	&$0.57\pm 0.05$	&$0.24\pm 0.04$		&$(3.3\pm 0.7)\times 10^{-3}$	&$0.21\pm 0.04$	\\
$50$	&$0.1$	&$1000$	&$92$	&$0.38\pm 0.03$	&$0.11\pm 0.02$		&$(1.2\pm 0.2)\times 10^{-3}$	&$0.15\pm 0.02$	\\
$50$	&$0.1$	&$3000$	&$95$	&$0.29\pm 0.03$	&$0.08\pm 0.02$		&$(5.2\pm 1.2)\times 10^{-4}$	&$0.13\pm 0.03$	\\
$50$	&$0.1$	&$10\,000$ &$81$	&$0.28\pm 0.03$	&$0.04\pm 0.006$	&$(2.8\pm 0.5)\times 10^{-4}$	&$0.17\pm 0.04$	\\
$50$	&$0.1$	&$30\,000$ &$122$	&$0.28\pm 0.03$	&$0.016\pm 0.002$	&$(1.1\pm 0.3)\times 10^{-4}$	&$0.09\pm 0.03$	\\
$50$	&$0.1$	&$100\,000$ &$344$	&$0.26\pm 0.000$ &$0.007\pm 0.001$	&$(4.7\pm 0.7)\times 10^{-5}$ &$0.05\pm 0.01$	\\
$50$	&$1$	&$3$		&$63$	&$1.39\pm 0.16$	&$0.60\pm 0.06$		& $(1.3\pm 0.2)\times 10^{-1}$			&$0.32\pm 0.05$	\\
$50$	&$1$	&$10$		&$70$	&$1.03\pm 0.1$	&$0.43\pm 0.05$		&$(7.8\pm1.2)\times 10^{-2}$	&$0.34\pm 0.03$	\\
$50$	&$1$	&$30$		&$97$	&$0.65\pm 0.08$	&$0.27\pm 0.05$		&$(3.7\pm 0.8)\times 10^{-2}$	&$0.24\pm 0.04$	\\
$50$	&$1$	&$100$		&$90$	&$0.38\pm 0.03$	&$0.12\pm 0.02$		&$(1.2\pm 0.2)\times 10^{-2}$	&$0.15\pm 0.03$	\\
$50$	&$1$	&$300$		&$97$	&$0.29\pm 0.04$	&$0.07\pm 0.02$		&$(5.2\pm 1.3)\times 10^{-3}$	&$0.13\pm 0.03$	\\
$50$	&$1$	&$1000$	&$68$	&$0.27\pm 0.03$	&$0.04\pm 0.003$		&$(2.5\pm 0.4)\times 10^{-3}$	&$0.14\pm 0.02$	\\
$50$	&$1$	&$3000$	&$99$	&$0.28\pm 0.03$	&$0.02\pm 0.003$		&$(1.2\pm 0.3)\times 10^{-3}$	&$0.10\pm 0.02$	\\
$50$	&$1$	&$100\,000$ &$474$	&$0.62\pm 0.4$	&$0.0008 \pm 0.0004$ &$(7.0\pm 2.9) \times 10^{-5}$	&$0.006\pm 0.003$ \\
      \end{tabular}
    \end{center}
  \caption{Summary of all the runs obtained using the standard equations, with parameters $Re$, $Pe$ and $B$. Quantities in columns 5--8 are computed in the manner described in section \ref{sec:dataex}.} 
  \label{tab:NormalTable}
  \end{table}

  \begin{table}
    \begin{center}
    	\setlength{\tabcolsep}{8pt}
      \begin{tabular}{cccccc}
 $Re$ & $BPe$ & $Re_{\lambda}$ & $l_z \pm \delta l_z$ & $w_{rms} \pm \delta w_{rms}$  & $\eta \pm \delta \eta$  \vspace{0.15cm} \\   
$600$	&$1$	&$236$	&$1.451\pm	0.184$ &$0.814\pm 0.084$ &$0.16\pm	0.04$\\
$600$	&$10$	&$246$	&$0.962\pm	0.074$ &$0.566\pm 0.045$ &$0.38\pm	0.07$\\
$600$	&$40$	&$276$	&$0.613\pm	0.028$ &$0.485\pm 0.041$ &$0.37\pm	0.05$\\
$600$	&$100$	&$266$	&$0.455\pm	0.029$ &$0.379\pm 0.037$ &$0.40\pm	0.05$\\
$600$	&$300$	&$281$	&$0.319\pm	0.012$ &$0.297\pm 0.038$ &$0.38\pm	0.04$\\
$600$	&$600$	&$383$	&$0.254\pm	0.016$ &$0.249\pm 0.041$ &$0.35\pm	0.03$ \vspace{0.15cm} \\

$300$	&$1$	&$166$	&$1.609\pm	0.191$ &$0.864\pm 0.104$ &$0.17\pm	0.04$\\
$300$	&$10$	&$167$	&$0.992\pm	0.093$ &$0.609\pm 0.042$ &$0.38\pm	0.04$\\
$300$	&$40$	&$202$	&$0.619\pm	0.028$ &$0.457\pm 0.036$ &$0.41\pm	0.04$\\
$300$	&$100$	&$262$	&$0.458\pm	0.020$ &$0.378\pm 0.048$ &$0.40\pm	0.03$\\
$300$	&$300$	&$427$	&$0.312\pm	0.015$ &$0.256\pm 0.030$ &$0.36\pm	0.02$\\
$300$	&$600$	&$182$	&$0.218\pm	0.000$ &$0.122\pm 0.005$ &$0.40\pm	0.02$ \vspace{0.15cm} \\

$100$	&$1$	&$98$	&$1.731\pm	0.376$&$	0.771\pm 0.059$ &$0.18\pm	0.04$\\
$100$	&$10$	&$96$	&$0.933\pm	0.064$&$	0.478\pm 0.036$ &$0.43\pm	0.04$\\
$100$	&$100$	&$184$	&$0.411\pm	0.029$&$	0.198\pm 0.035$ &$0.29\pm	0.04$\\
$100$	&$300$	&$87$	&$0.309\pm	0.008$&$	0.096\pm 0.007$  &$0.26\pm	0.03$\\
$100$	&$600$	&$87$	&$0.255\pm	0.011$&$	0.063\pm 0.006$ &$0.21\pm	0.03$\\
      \end{tabular}
    \end{center}
  \caption{Summary of all the runs obtained using the low P\'eclet number equations, with parameters $Re$, and $BPe$. Quantities in columns 3--5  are computed in the manner described in section \ref{sec:dataex}.} 
    \label{tab:LPNTable}
  \end{table}

Tables \ref{tab:NormalTable} and \ref{tab:LPNTable} present all the runs that have been performed with this model setup, using equations (\ref{eqn:momentumnondim})-(\ref{eqn:continuitynondim}) and (\ref{eqn:lpnequation1})-(\ref{eqn:lpnequation2}), respectively. To save on computational time, only one of the simulations at each Reynolds number is initiated from the original initial conditions (i.e. ${\bf u} = \sin (y) {\bf e}_x$ plus some small amplitude white noise). All the others are restarted from the end point of a simulation at nearby values of $Pe$ or $B$. In all cases, we have run the simulations until they reach a statistically stationary state, except where explicitly mentioned. Note that for very large values of $B$ or very small values of $Pr$, we have found it necessary to decrease the value of the maximum allowable timestep substantially. This is because the system of equations becomes increasingly stiff and is otherwise susceptible to the development of spurious elevator modes (i.e. modes that are invariant in the vertical direction). To save on computational time, we only ran simulations using the standard equations in that limit. 

The number of Fourier modes used in each direction (after dealiasing) depends on the Reynolds number $Re$ selected. In terms of equivalent grid points, the resolution used is 192$\times$96$\times$96 ($Re = 50$ runs), 384$\times$192$\times$192 ($Re = 100$ runs), 576$\times$288$\times$288 ($Re = 300$ runs) and 768$\times$384$\times$384 ($Re = 600$ runs). The same resolution is used regardless of the values of $B$ and $Pe$. We have verified that the product of the maximum wavenumber and the Kolmogorov scale is always greater than one (it is about 1.1 for the $Re = 600$ runs, and increases as $Re$ decreases).

\subsection{Typical simulations: early phase}

\begin{figure}
	\centering
	\vspace{0.25cm}
	\includegraphics[width=\linewidth]{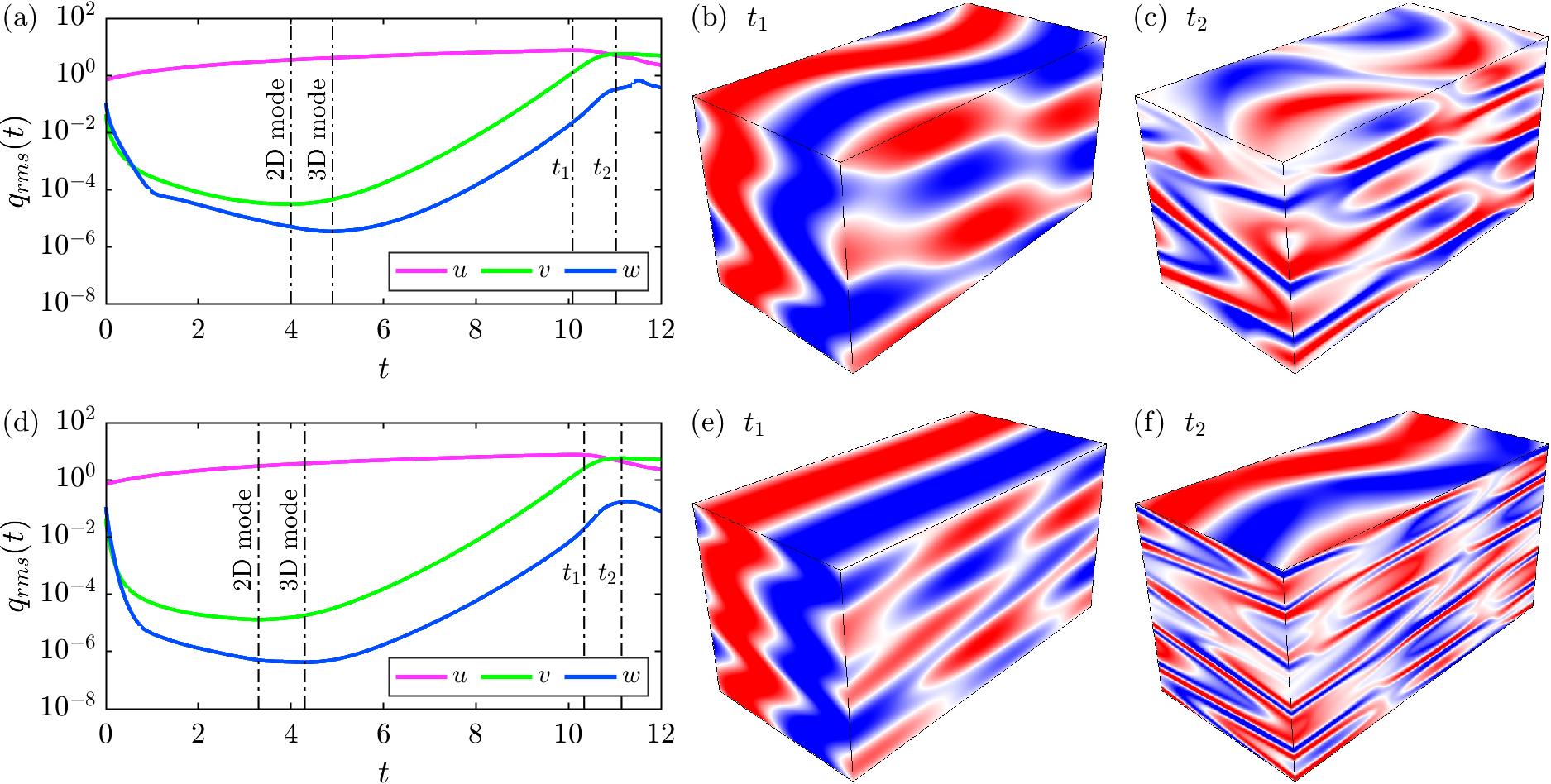}
	\caption{(a) Time evolution of the r.m.s. velocities in a simulation with $Re = 300$, $Pe = 0.1$ and $B = 30\,000$. The onset of the 2D modes ($k_z=0$) and 3D modes ($k_z \ne 0$) of instability are indicated. (b) and (c): Snapshots of the streamwise velocity at times $t_1$ and $t_2$ for the same simulation as panel (a). (d) As in (a), except with $B = 300\,000$.  (e) and (f): Snapshots of the streamwise velocity at times $t_1$ and $t_2$ for the same simulation as in panel (d). Note the change in the vertical scale as $B$ increases. \label{fig:earlysnaps}}
\end{figure}

We begin by presenting the early phases of development of the horizontal shear instability, in two typical simulations at moderately large Reynolds number ($Re = 300$), high stratification ($B = 30\,000$ and $300\,000$, respectively) and relatively low P\'eclet number ($Pe = 0.1$). Both simulations were initialized with ${\bf u} = \sin (y) {\bf e}_x$  plus small amplitude white noise. Figures \ref{fig:earlysnaps}(a) and \ref{fig:earlysnaps}(d) show the root mean square (r.m.s.) values of the streamwise ($u_{rms}$), spanwise $(v_{rms})$ and vertical $(w_{rms})$ velocities for each simulation, computed at each instant in time as 
\begin{equation}
q_{rms}(t) =  \langle q^2 \rangle^{1/2}, 
\label{eq:qrmsdef}
\end{equation}
where the angular brackets denote a volume average such that 
\begin{equation}
\langle q \rangle = \frac{1}{L_xL_yL_z} \int q(x,y,z,t) dxdydz .
\label{eq:qavdef}
\end{equation}
For both values of $B$, we clearly see the growth of the streamwise flow due to the forcing. Spanwise and vertical fluid motions first decay, until the onset of the 2D mode of instability (i.e. whereby $v_{rms}$ begins to grow while $w_{rms}$ continues to decay), rapidly followed by the 3D mode, for which $w_{rms}$ finally also begins to grow. 

Snapshots of the streamwise velocity fields near the saturation of these instabilities are presented in figures \ref{fig:earlysnaps}(b)-(c) and \ref{fig:earlysnaps}(e)-(f).  In both cases, the snapshot at $t_1$ illustrates the early development of the 2D and 3D modes of instability. The 2D mode causes a meandering of the background flow, and the 3D mode causes a vertical modulation of the position of the meanders. We also see that the 3D mode has a substantially smaller vertical scale for larger $B$. The snapshot at $t_2$ shows how the instability further evolves with time: the meanders and their vertical shifts both grow in amplitude, leading to the development of substantial vertical shear of the streamwise flow. 

While similar early-time dynamics are observed at all parameter values (assuming the 2D mode is unstable), what happens beyond that depends on $Re$, $B$ and $Pe$. We now describe in turn the various regimes that can be found. 

\subsection{Typical simulations: nonlinear saturation} \label{subsec:nlsaturation}

\begin{figure}
	\centering
	\vspace{0.25cm}
	\includegraphics[width=1\linewidth]{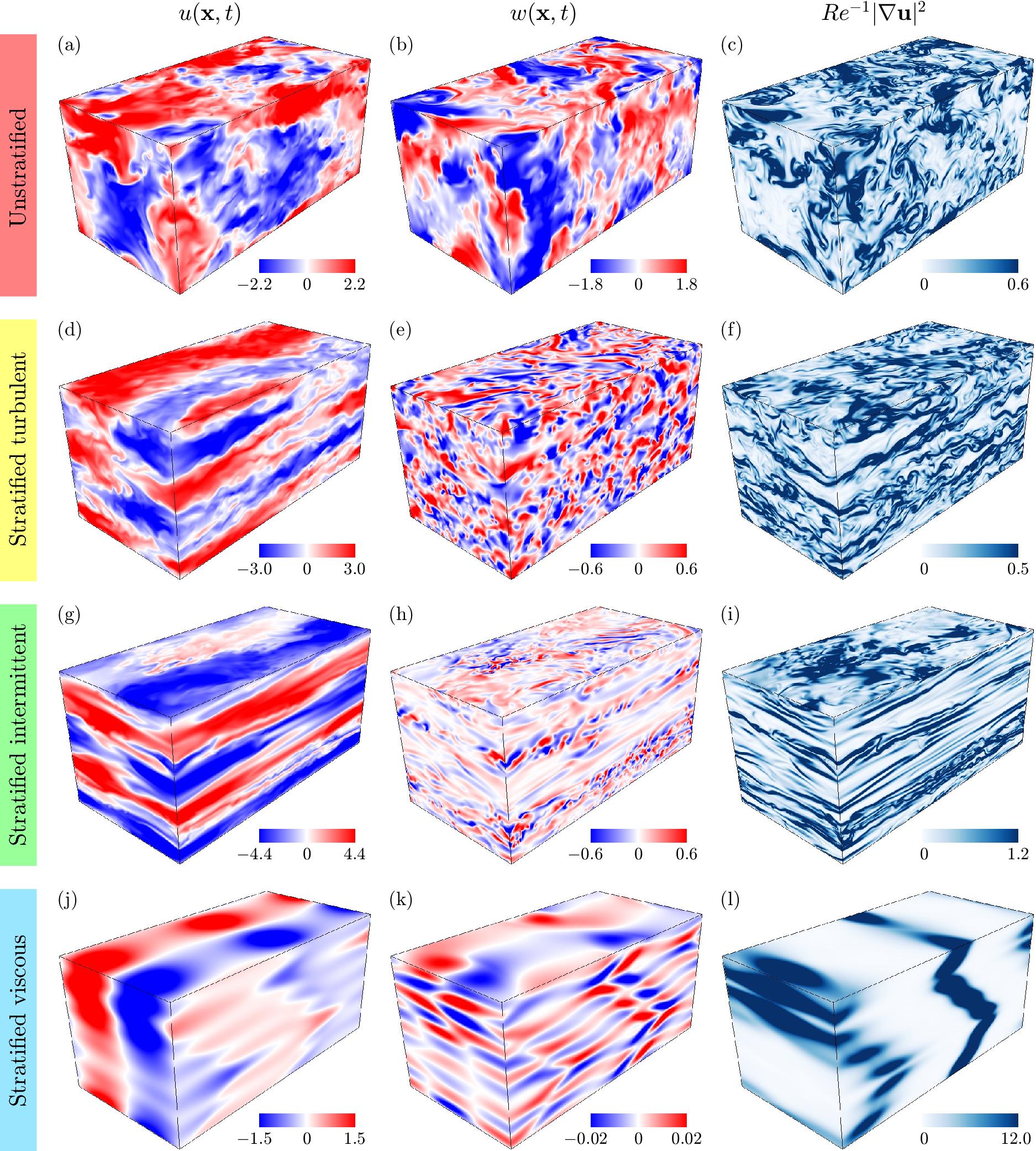}
	\caption{Snapshots of the streamwise velocity (left), vertical velocity (centre) and local viscous dissipation rate (right) during the statistically stationary states of DNSs with $Pe=0.1$ and: (a-c) $Re=300$, $B=1$; (d-f) $Re=300$, $B=100$; (g-i) $Re=300$, $B=10\,000$; (j-l) $Re=50$, $B=100\,000$. Each of these examples are characteristic of a particular regime, listed on the left.    
		\label{fig:samplesnaps}}
\end{figure}

The nonlinear saturation of this body-forced horizontal shear flow depends crucially on the selected value of the stratification parameter $B$. In what follows, we investigate the effect of varying $B$. Snapshots of the streamwise velocity, vertical velocity and local viscous dissipation rate, taken during the statistically stationary state, are presented in figure \ref{fig:samplesnaps}. In all but the last row, $Re = 300$, and $Pe = 0.1$. For the last row, $Re = 50$.

For very large values of $B$ (bottom row in figure \ref{fig:samplesnaps}), the vertical scale of the 3D mode of instability is relatively small. Even though substantial shear develops between successive meanders of the streamwise jets, this shear is too small to overcome the stabilizing effect of viscosity, and remains stable. The resulting flow takes the form of thin layers, crucially in the \emph{velocity} field, each of which presents a meandering jet with its own distinct phase. These jets are weakly coupled in the vertical direction through viscosity. The vertical velocity field is small but non-zero, however, and is presumably generated by the weak horizontal divergence of the flow within each jet. 

As $B$ decreases (i.e. moving up in figure \ref{fig:samplesnaps}), the reduced stratification now allows for the development of secondary vertical shear instabilities between the meanders, albeit only intermittently, with correspondingly larger vertical velocities. Spatially localized overturns can be seen in figures \ref{fig:samplesnaps}(g)-(i). These become more numerous and more frequent as $B$ continues to decrease. The viscous dissipation is clearly enhanced in the turbulent regions compared with the laminar regions. 

For intermediate values of $B$ (see figures \ref{fig:samplesnaps}(d)-(f)), the flow becomes fully turbulent. The vertical scale of the eddies remains relatively small, however, consistent with stratification playing a role in shaping the dynamics of the turbulence. The meandering streamwise jets are still clearly visible. The dissipation rate snapshot shows that the scale of the turbulent eddies is small in both horizontal and vertical directions. 

Finally, for low values of $B$, the scale of the eddies is now the domain scale, and the turbulence is unaffected by stratification. In fact, this system is very similar to the one obtained in weakly stratified vertically sheared flows \citep[see][]{GaraudKulen16}, except for the horizontally-averaged mean flow (which varies with $y$ instead of $z$). 

These observations therefore suggest the existence of at least four distinct LPN dynamical regimes: unstratified turbulence for very low $B$; stratified turbulence for intermediate values of $B$; intermittent turbulence for higher values of $B$; and finally, viscously-dominated stratified laminar flow for the highest values of $B$. We will now proceed to characterize these different regimes more quantitatively.

\subsection{Data extraction}
\label{sec:dataex}

Each of the simulations we have performed was integrated until the system reached a statistically stationary state. This can take a long time, especially for the very strongly stratified systems, so data in that limit is scarce except for the lowest values of $Re$. Once in that statistically stationary state, we compute the time average, and deviations around that average, of $w_{rms}(t)$ and $T'_{rms}(t)$, where $q_{rms}(t)$ for any quantity $q$ was defined in (\ref{eq:qrmsdef}). These are reported as $w_{rms} \pm \delta w_{rms}$ and $T'_{rms} \pm \delta T'_{rms}$, respectively, in tables \ref{tab:NormalTable} and \ref{tab:LPNTable}.

We also compute the temperature flux as
\begin{equation}
F_T(t) = \langle wT' \rangle
\end{equation}
for DNSs that use the standard equations and 
\begin{equation}
Pe^{-1} F_T(t) = \langle w \nabla^{-2} w  \rangle 
\end{equation}
for DNSs that use the LPN equations, where the angular bracket was defined in (\ref{eq:qavdef}). We finally compute the viscous energy dissipation rate as
\begin{equation}
\epsilon(t) = Re^{-1} \langle  |\nabla {\bf u} |^2 \rangle,
\end{equation}
where $\epsilon$ is the non-dimensional version of $\varepsilon$ introduced in section \ref{sec:intro2}. Even though the turbulence is highly anisotropic, we can use $\epsilon$ to compute the Reynolds number based on the Taylor microscale, which in our non-dimensionalization is given by 
	\begin{equation}
		Re_\lambda = \sqrt{\frac{15 Re}{\epsilon}} \, U_{rms}^2,
	\end{equation}
	where $U_{rms}$ is the total r.m.s. velocity defined as 
	\begin{equation}
		U_{rms} = \left( \frac{1}{3} \left( u^2_{rms} + v^2_{rms} + w^2_{rms} \right) \right)^{1/2}.
	\end{equation}
The quantity $Re_{\lambda}$, whose interpretation for homogeneous isotropic turbulence is well documented \citep{taylor1935,pope2000,davidson2015}, is reported in tables 1 and 2, and reaches values between 250 and 500 for the $Re = 600$ simulations.

\begin{figure}
	\centering
	\vspace{0.25cm}
	\includegraphics[width=\linewidth]{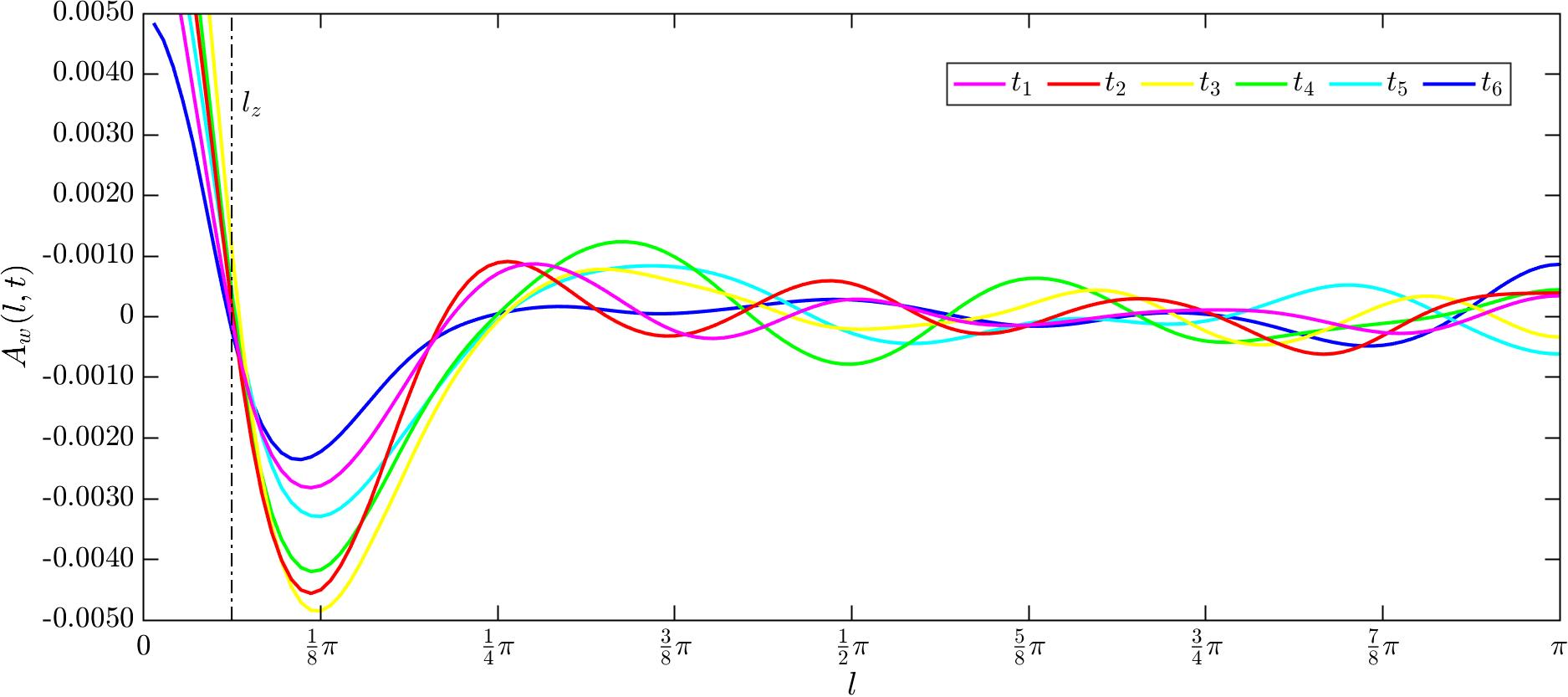}
	\caption{Autocorrelation function $A_w(l,t)$ as defined in (\ref{eq:autocorel}) computed at six randomly selected times during the statistically stationary state of a simulation with parameters $Re=300$, $B=10\,000$ and $Pe=0.1$. Note how  $A_w(l,t)$ has a well-defined first zero, whose time-average defines the vertical eddy scale $l_z$. \label{fig:autocorel}}
\end{figure}

We can also use the data to diagnose the dominant energetic balance taking place in the system. Indeed, dotting the momentum equation (\ref{eqn:momentumnondim}) with ${\bf u}$ and integrating over the domain, we obtain
\begin{eqnarray}
\frac{\partial}{\partial t} \langle \frac{1}{2} | {\bf u} |^2 \rangle &=& B \langle wT' \rangle - \frac{1}{Re} \langle | \nabla {\bf u} |^2 \rangle  +  \langle u \sin (y) \rangle ,   \\
&=& B F_T  - \epsilon +  \langle u \sin (y) \rangle .
\end{eqnarray} 
This shows that the rate at which the body force does work on the flow, $\langle u \sin (y) \rangle$, is partitioned between energy that is dissipated viscously (through $\epsilon$), and energy that is converted into potential energy (at a rate $BF_T$). The fate of the latter can be established by multiplying the temperature equation by $T'$ and integrating over the domain, which reveals that 
\begin{equation}
\frac{\partial}{\partial t} \langle \frac{1}{2} T'^2 \rangle  + B F_T =  \langle \frac{1}{Pe} T' \nabla^2 T' \rangle  = - \langle \frac{1}{Pe} |\nabla T'|^2 \rangle 
\end{equation} 
for the full equations (while in the LPN limit, the time derivative simply disappears). This shows that $BF_T$ is ultimately dissipated thermally at a rate $Pe^{-1} \langle | \nabla T' |^2\rangle$.  

From these considerations, it is common to define a so-called instantaneous mixing efficiency \citep[see e.g.][]{Maffioli2016}
\begin{equation}
\eta(t) = \frac{-BF_T(t) }{-BF_T(t) + \epsilon(t)}  = \frac{-BF_T(t) }{\langle u \sin (y) \rangle} 
\end{equation}
at a given point in time, which measures the efficiency with which kinetic  energy, produced by the applied forcing, is converted into potential energy as opposed to being dissipated viscously. 
We have computed $\eta(t)$ for all simulations produced, and report its time average and deviation from that average, while in a statistically stationary state, as $\eta \pm \delta \eta$ in tables \ref{tab:NormalTable} and \ref{tab:LPNTable}. 

Finally, another useful diagnostic of the flow is the typical vertical scale of the turbulent eddies. As discussed in \citet{GaraudKulen16} and \citet{Garaudal17}, there are many different ways of extracting such a length scale, either from weighted averages over the turbulent  energy spectrum, or from spatial autocorrelation functions of the velocity field. \citet{Garaudal17} compared these different methods and concluded that the spatial autocorrelation function was a more physical and reliable way of extracting the vertical length scale. In what follows, we therefore compute the function
\begin{equation}
A_w(l,t) = \frac{1}{L_xL_yL_z} \int w(x,y,z,t) w(x,y,z+l,t) dxdydz
\label{eq:autocorel}
\end{equation}
at each timestep for which the full fields are available, using periodicity of $w$ to deal with points near the domain boundaries. Sample functions for six randomly selected times during the statistically stationary state are shown in figure \ref{fig:autocorel} for a simulation with parameters $Re=300$, $B=10\,000$ and $Pe=0.1$ (a simulation from the stratified intermittent regime, snapshots from which are shown in figures \ref{fig:samplesnaps}(g)-(i)). We clearly see that $A_w(l,t)$ has a well-defined first zero at each timestep, which we call $l_z(t)$. The vertical eddy scale thus obtained is then averaged over all available timesteps during the statistically stationary state to obtain the mean vertical eddy scale $l_z$ and its standard deviation $\delta l_z$.


\section{Nonlinear saturation: scaling regimes} 
\label{sec:nonlinearregimes}

In our quest for a quantitative description of the four dynamical regimes described in section \ref{subsec:nlsaturation}, we endeavour to derive scaling laws that explain our data, in an analogous fashion to the approach of \citet{Brethouwer2007} in which the focus was on geophysically-relevant parameters ($Pr \gtrsim O(1)$). Consistent with our goal to study systems in which the P\'{e}clet number $Pe$ is small, we have run a range of simulations using the standard equations (\ref{eqn:momentumnondim})-(\ref{eqn:continuitynondim}) for three different P\'{e}clet numbers (0.01, 0.1 and 1), which we compare alongside simulations using the LPN equations (\ref{eqn:lpnequation1}) and (\ref{eqn:lpnequation2}), noting excellent agreement. Using both sets of equations, we consider four different Reynolds numbers (50, 100, 300, 600) and investigate a wide range of background stratifications. 

\subsection{Effects of stratification on mixing and the vertical scale of eddies}

\begin{figure}
	\centering
	\vspace{0.25cm}
	\includegraphics[width=\linewidth]{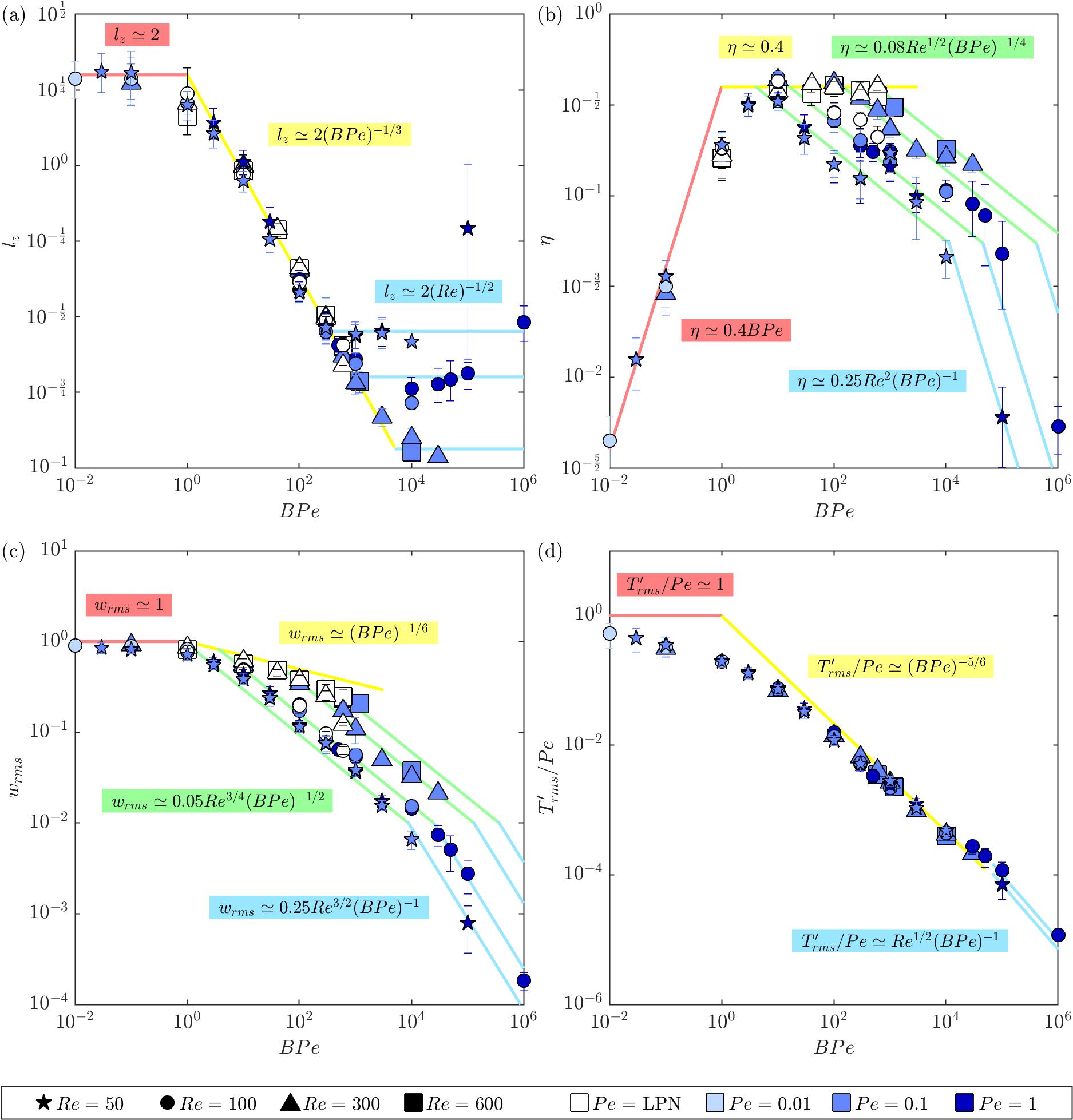}
	\caption{Variation with $BPe$ of four diagnostics, defined in section \ref{sec:dataex}: (a) $l_z$, (b) $\eta$, (c) $w_{rms}$, (d) $T^{\prime}_{rms}/Pe$. All DNSs listed in tables \ref{tab:NormalTable} and \ref{tab:LPNTable} are plotted, with shapes indicating the Reynolds number and colours indicating the P\'eclet number. Coloured lines illustrate our proposed scalings for the (red) unstratified, (yellow) stratified turbulent, (green) stratified intermittent and (blue) stratified viscous regimes.}
	\label{fig:dnsscalings}
\end{figure}

The first flow diagnostic that we discuss is the vertical eddy scale $l_z$, computed using the method described in section \ref{sec:dataex}. Figure \ref{fig:dnsscalings}(a) shows $l_z$ as a function of $BPe$, consistent with our expectations on the potential relevance of this parameter for low P\'eclet number flows (as discussed in section \ref{subsec:lpnmodel}). For all but the largest values of $BPe$ (which corresponds to the viscous regime discussed in section \ref{subsec:nlsaturation}), we confirm that $BPe$ is indeed the relevant parameter, and that $l_z$ is independent of $Re$. As a result, all the data collapses on a single universal curve.
For weak stratification, which we refer to as the unstratified regime, the vertical eddy scale is invariant with respect to both stratification and Reynolds number. We find that $l_z \simeq 2$, which is of the order of the size of the periodic domain. For intermediate values of $BPe$, corresponding to the stratified turbulent regime described in section \ref{subsec:nlsaturation}, we find that
\begin{equation}
l_z \simeq 2(BPe)^{-1/3}
\end{equation}
with some uncertainty in both the prefactor and the exponent due to the inherent variability of the flow.

Finally, for very strong stratification in the stratified viscous regime, the vertical eddy scale appears to become independent of $BPe$ and now depends solely on Reynolds number, with the empirical relationship given by:
\begin{equation} \label{eqn:unstratifiedlz}
l_z \simeq 2Re^{-1/2},
\end{equation}
again with some uncertainty in the scaling and exponent. This scaling is analogous to the viscously-affected stratified regime considered by \citet{Brethouwer2007} and discussed in the introduction (since $l_z$ in (\ref{eqn:unstratifiedlz}) is non-dimensional and scaled by a characteristic horizontal length scale). While only three clear regimes are evident in this plot, data from the stratified intermittent regime discussed in section \ref{subsec:nlsaturation} lies in the region of parameter space between the $l_z \sim (BPe)^{-1/3}$ and $l_z \sim Re^{-1/2}$ regimes, as the DNSs begin to feel the effects of viscosity, and hence $Re$.

It is also of interest to observe how the mixing efficiency $\eta$, discussed in section \ref{sec:dataex}, depends on the stratification $BPe$ and Reynolds number $Re$. Figure \ref{fig:dnsscalings}(b) shows $\eta$ as a function of $BPe$ for each of our simulations. This time, the four regimes can be clearly identified. For the unstratified regime, the mixing efficiency depends only on $BPe$, and is given by:
\begin{equation}
\eta \simeq 0.4 BPe.
\end{equation}
As the stratification increases, the mixing efficiency increases until it reaches a plateau at $\eta \simeq 0.4$ which, as we argue below, is a defining property of the stratified turbulent regime. The range of values of $BPe$ for which $\eta$ is approximately constant is very small for $Re = 50$, but clearly increases with $Re$, and is quite substantial for $Re = 600$. However, in all cases, a threshold is reached where $\eta$ begins to decrease again. To understand why this is the case, note that the vertical eddy scale decreases rapidly (as discussed above) as the stratification increases and inevitably reaches a point where the effects of viscosity begin to play a role. This is manifest in the fact that $\eta$ begins to depends on $Re$. The system enters the intermittently turbulent regime, where we observe the new empirical scaling:
\begin{equation} \label{eq:etaintermittent}
\eta \simeq 0.08 Re^{1/2}(BPe)^{-1/4},
\end{equation}
with significant uncertainty in the scalings owing to the high variability of $\eta$ in this  regime. Finally, for even larger values of $BPe$, our DNSs suggest a fourth regime for very large stratification, where we tentatively observe the scaling
\begin{equation}
\eta \simeq 0.25 Re^{2}(BPe)^{-1}
\end{equation}
which we described earlier (see section \ref{subsec:nlsaturation}) as characteristic of the stratified viscous regime. Note that the observed scaling in this regime is the most uncertain, as very little data is available.

Analogous empirical scalings are evident in figures \ref{fig:dnsscalings}(c) and \ref{fig:dnsscalings}(d) for the respective variations with $BPe$ of the vertical velocity field $w_{rms}$ and the temperature perturbation field $T^{\prime}_{rms} / Pe$. These observations inspire us to attempt to derive scaling laws using ideas of dominant balance in the governing equations.

\subsection{Derivation of scaling regimes} \label{subsec:nonlinearscalingregimes}

In the following analysis, and consistent with our study of low P\'eclet number systems, we always assume a LPN balance in equation (\ref{eqn:densitynondim}) such that 
\begin{equation} \label{eqn:regime_LPN}
w \simeq \frac{1}{Pe} \nabla^2 T^{\prime}.
\end{equation}
Our approach, therefore, is to consider the dominant balance between terms in the momentum equation (\ref{eqn:momentumnondim}), specifically the relative importance of stratification, inertia and viscosity.

\subsubsection{Unstratified regime}

We begin by considering the unstratified regime, described in section \ref{subsec:nlsaturation} and illustrated in figures \ref{fig:samplesnaps}(a)-(c). Motivated by the qualitative observation of the domain-filling eddies in figures \ref{fig:samplesnaps}(a) and \ref{fig:samplesnaps}(b), we make the assumptions that each of the three velocity components and eddy length scales are approximately isotropic with
\begin{equation} \label{eqn:unstratifiedassumptions}
u_{rms}, v_{rms}, w_{rms} \sim O(1); \quad l_x, l_y, l_z \sim O(1).
\end{equation}
These assumptions for $w_{rms}$ and $l_z$ are confirmed in figures \ref{fig:dnsscalings}(a) and \ref{fig:dnsscalings}(c), indicated by the red lines. By combining the LPN approximation (\ref{eqn:regime_LPN}) with assumptions (\ref{eqn:unstratifiedassumptions}), we find a scaling for the typical temperature perturbations:
\begin{equation} \label{eqn:regime1_T}
\frac{T^{\prime}_{rms}}{Pe} \sim O(1).
\end{equation}
In terms of the mixing efficiency $\eta$, 
(\ref{eqn:unstratifiedassumptions}) implies $\langle u \sin (y) \rangle \sim O(1)$. Thus 
\begin{equation} \label{eqn:regime1_eta}
\eta \sim \frac{B\langle wT^{\prime} \rangle}{\langle u \sin (y) \rangle} \sim B w_{rms} T^{\prime}_{rms}  \sim BPe. 
\end{equation} 
The theoretically derived scalings (\ref{eqn:regime1_T}) and (\ref{eqn:regime1_eta}) are consistent with the empirical scalings determined using our DNSs, shown using the red lines in figures \ref{fig:dnsscalings}(d) and \ref{fig:dnsscalings}(b) respectively. The lack of $Re$-dependence affirms the irrelevance of viscosity in this regime.

Finally, it is of interest to compute the condition of validity for this unstratified regime. In the vertical momentum equation (\ref{eqn:momentumnondim}), we have assumed that stratification is weak relative to fluid inertia, such that $BT^{\prime} \ll \mathbf{u} \cdot \nabla w$. Using the scalings derived above, we find that this is true when
\begin{equation} \label{eqn:regime1_condition}
BPe \ll O(1).
\end{equation}
Condition (\ref{eqn:regime1_condition}) combined with the condition for linear instability ($Re>2^{1/4}$) defines the region of parameter space in which we would expect to observe this regime of unstratified turbulence.

\subsubsection{Stratified turbulent regime}
\label{sec:stratreg}

As the stratification increases, the system transitions into the stratified turbulent regime, first presented in section \ref{subsec:nlsaturation}. This regime is defined by a constant mixing efficiency. Inspection of the snapshots in figures \ref{fig:samplesnaps}(d) and \ref{fig:samplesnaps}(e) reveals that the vertical velocity field, which is generated by localized shear-driven Kelvin-Helmholtz type instabilities, is mostly small-scale. By constrast, the horizontal velocity field contains both large scales (the modulated meanders) and small scales (associated with the small vertical scales). Consequently, we assume that
\begin{equation} \label{eqn:turbassumptions}
u_{rms}, v_{rms} \sim O(1); \quad l_x \sim l_y \sim l_z.
\end{equation}
With this assumption, the LPN approximation becomes
\begin{equation}
w_{rms} \sim  Pe^{-1}  \frac{T^{\prime}_{rms} }{ l_z^2} . \label{eq:LPNrms}
\end{equation}
Since the vertical flow is generated by shear instabilities of the horizontal flow, and since stratification is now important, we anticipate that the dominant balance in the vertical momentum equation should be $\mathbf{u} \cdot \nabla w \sim BT^{\prime}$, implying that 
\begin{equation} \label{eqn:turbvertbalance}
u_{rms} w_{rms} l_z^{-1} \sim BT^{\prime}_{rms}. 
\end{equation}
We note that this implicitly assumes that the vertical pressure gradient $\partial p / \partial z$ is either of the same order as $BT^{\prime}_{rms}$ or much smaller, which on the surface appears to contradict the fact that $p$ ought to be $O(1)$ based on the horizontal component of the momentum equation. However, the contradiction can be resolved by noting that $p$, similar to $u$ and $v$, has both a large-scale and a small scale component, and that only the large-scale component is $O(1)$ while it is the small-scale component (of unknown amplitude) that mostly contributes to the vertical derivative. While a rigorous multi-scale analysis (which is beyond the scope of this paper) will be required to formalize this argument, we note that since both the inertial and the buoyancy terms must play a role in the dynamics of the flow, the $\partial p / \partial z$ term cannot replace either $\mathbf{u} \cdot \nabla w$ or $BT^{\prime}$ in the dominant balance (at best, it can be of the same order of magnitude). Combining with  (\ref{eq:LPNrms}) and $u_{rms} \sim O(1)$ leads to the vertical eddy length scale
\begin{equation} \label{eqn:turblzscaling}
l_z \sim (BPe)^{-1/3},
\end{equation}
which is confirmed by the yellow line in figure \ref{fig:dnsscalings}(a). Empirically, we find that the prefactor is close to 2, and confirm that this scaling is independent of the Reynolds number. 

As mentioned earlier, $\eta \sim O(1)$ is a defining property of the stratified turbulent regime, with a roughly equal partitioning between viscous dissipation and thermal dissipation. The yellow line in figure \ref{fig:dnsscalings}(b) suggests that this constant value of the mixing efficiency is
\begin{equation}
\eta \simeq 0.4.
\end{equation}
Since $\langle u \sin (y) \rangle \sim O(1)$ from assumption (\ref{eqn:turbassumptions}), then $\eta \sim B\langle wT^{\prime} \rangle$ implies that 
\begin{equation} \label{eqn:turbeta}
B w_{rms}T_{rms}^{\prime}  \sim O(1).
\end{equation}
Combining (\ref{eqn:turbeta}) with the LPN approximation (\ref{eqn:regime_LPN}) and the vertical momentum equation balance (\ref{eqn:turbvertbalance}) leads to the additional scalings
\begin{equation} \label{eqn:turbwTscalings}
\frac{T'_{rms}}{Pe} \sim (BPe)^{-5/6}; \quad w_{rms} \sim (BPe)^{-1/6}.
\end{equation}
There is strong evidence for both of these scalings as illustrated by the yellow lines in figures \ref{fig:dnsscalings}(d) and \ref{fig:dnsscalings}(c) respectively, and the empirical data are consistent with the associated prefactors being close to one in each case. Once again, we highlight the lack of dependence on $Re$ in (\ref{eqn:turbwTscalings}).

In this regime, we can finally estimate a generic non-dimensional turbulent diffusivity for vertical transport of a passive scalar as
\begin{equation}\label{eq:strat_turb_diff}
D_{\rm turb} \sim w_{rms} l_z \sim (BPe)^{-1/2}, 
\end{equation}
with a prefactor that is expected to be of order unity. This result can be compared with the mixing coefficient expected in low P\'eclet number stratified turbulence caused by vertical shear, which scales as $(RiPe)^{-1}$ instead (see (\ref{eq:Dturb}), when cast in non-dimensional form). We see that $D_{\rm turb}$ decreases much less rapidly with increasing stratification in horizontally-sheared flows than in vertically sheared flows, at least while the system is in this stratified turbulent regime.

The assumptions that we made in the vertical momentum equation balance, i.e. that the viscous terms are negligible ($Re^{-1}\nabla^2 w \ll BT^{\prime}$), along with scalings for $l_z$, $u_{rms}$ and $T^{\prime}_{rms}$, lead to the condition $BPe \ll Re^2$. This suggests that the stratified turbulent regime scalings should apply when:
\begin{equation} \label{eqn:regime2_condition}
1 \ll BPe \ll Re^2.
\end{equation}
Condition (\ref{eqn:regime2_condition}), computed more precisely in section \ref{subsec:stratifiedintermittent}, uniquely defines the region of parameter space in which we would expect to observe this particular type of stratified turbulence in flows at low P\'eclet number.

\subsubsection{Stratified viscous regime}

As discussed in section \ref{subsec:nlsaturation}, for very strong stratification we observe the formation of thin and viscously coupled layers, each containing almost two-dimensional flow. Consequently, we expect that horizontal and vertical velocity components and length scales will both be strongly anisotropic. Denoting horizontal length scales as $l_h$, we make the following assumptions:
\begin{equation}
l_h \sim O(1); \quad l_z \ll l_h;
\end{equation}
\begin{equation}
u_{rms}, v_{rms} \sim O(1); \quad w_{rms} \ll u_{rms}, v_{rms}.
\end{equation}

In what follows, we split the velocity field into a horizontal and vertical component, $\mathbf{u} = \mathbf{u}_h + w \mathbf{e}_z$, with a corresponding decomposition of the gradient operator $\nabla = (\nabla_h, \partial / \partial z)$. The momentum equation can be split into its horizontal and vertical components as:
\begin{equation} \label{eqn:horizontalmomeqn1}
\frac{\partial \mathbf{u}_h}{\partial t} + \mathbf{u}_h \cdot \nabla_h \mathbf{u}_h + w \frac{\partial \mathbf{u}_h}{\partial z} + \nabla_h p =  \frac{1}{Re}\left(\nabla_h^2 \mathbf{u}_h + \frac{\partial^2 \mathbf{u}_h}{\partial z^2} \right) + \sin (y) \mathbf{e}_x, 
\end{equation}
\begin{equation} \label{eqn:horizontalmomeqn2}
\frac{\partial w}{\partial t} + \mathbf{u}_h \cdot \nabla_h w + w \frac{\partial w}{\partial z} + \frac{\partial p}{\partial z} =  \frac{1}{Re}\left(\nabla_h^2 w + \frac{\partial^2 w}{\partial z^2} \right) + BT'.
\end{equation}
If we assume a dominant balance between viscosity and the forcing in the horizontal momentum equation (\ref{eqn:horizontalmomeqn1}), then $Re^{-1} \partial_z^2 \mathbf{u}_h \sim \sin (y) \mathbf{e}_x \sim O(1)$. This balance, combined with $\mathbf{u}_h \sim O(1)$, leads to the classical viscous scaling for the vertical length scales \citep[cf.][]{Brethouwer2007}:
\begin{equation} \label{eqn:regime4_lz}
l_z \sim Re^{-1/2}.
\end{equation}
Substantial evidence for this scaling is visible in figure \ref{fig:dnsscalings}(a), where the series of blue lines correspond to $l_z \simeq 2 Re^{-1/2}$ for each individual Reynolds number. Note that these strongly stratified DNSs exhibit large amplitude quasi-time-periodic behaviour, a feature that we believe to be an intrinsic property of such flows. We consequently attribute the large error bars associated with some DNSs to this observation.

In the vertical momentum equation, we assume that the dynamics are hydrostatic, therefore $\partial_z p \sim BT^{\prime}$ implies $p l_z^{-1} \sim BT^{\prime}_{rms}$. This approximation, combined with the requirement from the balance in the horizontal momentum equation that $p \sim O(1)$, and with the scaling (\ref{eqn:regime4_lz}) for $l_z$, gives us a scaling for temperature perturbations:
\begin{equation} \label{eqn:regime4_T}
\frac{T^{\prime}_{rms}}{Pe} \sim Re^{1/2}(BPe)^{-1}.
\end{equation}
This stratified viscous regime is considerably more challenging to simulate than the other three regimes, a consequence of the very small time steps required and long integration times. However, we see in figure \ref{fig:dnsscalings}(d) that the blue lines, which represent the scalings in (\ref{eqn:regime4_T}), fit the few available data points well, once again with a prefactor close to one. 

The LPN approximation (\ref{eq:LPNrms}), combined with (\ref{eqn:regime4_lz}) and (\ref{eqn:regime4_T}), leads to a scaling for the vertical velocity field:
\begin{equation}
w_{rms} \sim Re^{3/2} (BPe)^{-1}.
\end{equation}
Again we see a good correspondence between the blue curves in figure \ref{fig:dnsscalings}(c), which represent this scaling, and the data, with a prefactor of 0.25.

Using these results, we finally find that
\begin{equation}
\eta \sim \frac{B\langle wT' \rangle}{\langle u \sin (y) \rangle } \sim B w_{rms} T^{\prime}_{rms} \sim Re^2 (BPe)^{-1},
\end{equation}
with a prefactor of 0.25 for consistency with the data obtained for $w_{rms}$ and $T^{\prime}_{rms}$. 
This is consistent with observations for $Re=50$ and $Re=100$ in figure \ref{fig:dnsscalings}(b).  We can also estimate a generic non-dimensional turbulent diffusivity for vertical transport of a passive scalar as
\begin{equation}
D_{\rm turb} \sim w_{rms} l_z \sim Re(BPe)^{-1}  \sim (BPr)^{-1}
\end{equation}
with a prefactor that is again expected to be of order unity. 

The viscous regime is achieved in the opposite limit to the one derived in (\ref{eqn:regime2_condition}) for the stratified turbulent regime, namely when $BPe \gg Re^2$. 
Thus we find that the system parameters must satisfy 
\begin{equation} \label{eqn:regime4_condition}
2^{1/2} < Re^2 \ll BPe
\end{equation}
when combined with the condition for linear instability. Condition (\ref{eqn:regime4_condition}), computed more precisely in section \ref{sec:regimes}, defines the region of parameter space in which we would expect to observe this stratified viscous regime. We note for consistency that each of the scalings obtained here do depend on the value of the Reynolds number, as one would expect.

\subsubsection{Stratified intermittent regime} \label{subsec:stratifiedintermittent}

There exists a fourth regime, visible both in the DNSs and the results presented in figures \ref{fig:dnsscalings}(a)-(d). This final regime is a transitional regime that occurs between the stratified turbulent regime and the stratified viscous regime. As discussed in section \ref{subsec:nlsaturation}, it is inherently intermittent in the sense that we observe spatially and temporally localised patches of small-scale turbulence generated via vertical shear instabilities, surrounded by more laminar, viscously dominated flow. Whilst we have been unable to derive satisfactory scalings for this regime, we can nevertheless deduce some of them empirically from figures \ref{fig:dnsscalings}(b) and \ref{fig:dnsscalings}(c).

For instance, we see in figure \ref{fig:dnsscalings}(b) that the onset of this stratified intermittent regime (indicated by the green lines) is characterised by a sudden change in the dependence of the mixing efficiency $\eta$ on $BPe$, from the constant value of 0.4 observed in the stratified turbulent regime to a regime where $\eta$ is given by (\ref{eq:etaintermittent}). It is interesting and perhaps reassuring to note that the parameter group $BPe / Re^2$, which controls $\eta$ in this regime, is the same parameter group that appears in the viscous regime. Note that for $\eta \simeq 0.1$, we observe a temporary flattening of this scaling just before the onset of the viscous regime. It is certainly possible that this feature is an artefact of inherent variability in the simulations (and therefore the measurement of $\eta$ has larger associated error bars). It is interesting to note, however, that this ``knee" in the curve does occur for flows with at least three different Reynolds numbers.

In addition, figure \ref{fig:dnsscalings}(c) suggests that $w_{rms}$ scales as 
\begin{equation}
w_{rms} \simeq 0.05 Re^{3/4}(BPe)^{-1/2}.
\end{equation}
No clear scalings for $l_z$  or $T'_{rms}/Pe$ appear to be deducible from the numerical results.

\begin{figure}
	\centering
	\vspace{0.25cm}
	\includegraphics[width=0.72\linewidth]{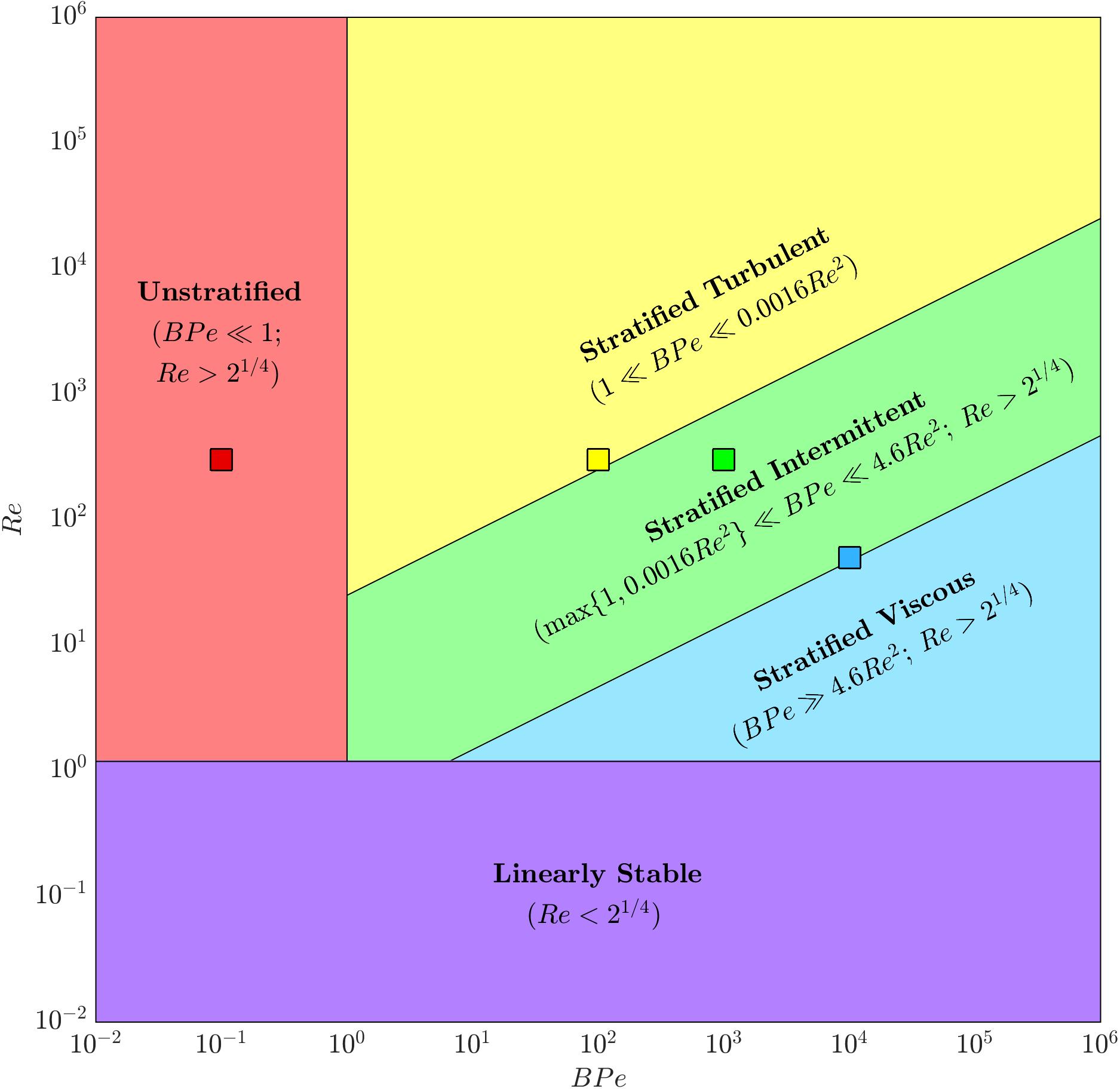}
	\caption{Regime diagram, applicable in the LPN limit, illustrating five dynamical regimes across system parameters $BPe$ (horizontal axis) and $Re$ (vertical axis). Each regime is associated with a colour: linearly stable (purple); unstratified (red); stratified turbulent (yellow); stratified intermittent (green); stratified viscous regime (blue). The four example DNSs presented in figure \ref{fig:samplesnaps} are associated with parameters corresponding to the red, yellow, green and blue squares.}
	\label{fig:dnsparameterspace}
\end{figure}

\subsection{Regime diagram}
\label{sec:regimes}

Figure \ref{fig:dnsparameterspace} summarises the four regimes of nonlinear saturation that were described in section \ref{subsec:nonlinearscalingregimes} along with the inclusion of the linearly stable regime that was discussed in section \ref{sec:linearstability}. The unstratified regime, indicated in red in figure \ref{fig:dnsparameterspace}, occurs when
\begin{equation}
BPe \ll O(1); \quad Re>2^{1/4}.
\end{equation}
The stratified turbulent regime, indicated by the yellow region in figure \ref{fig:dnsparameterspace}, and the stratified viscous regime, indicated by the blue region, exist when $BPe \ll Re^2$ and $BPe \gg Re^2$ respectively, with the stratified intermittent regime (green) lying at the transition. Written in terms of the buoyancy Reynolds number $Re_b = Re / B$, which is a key parameter identified by \cite{Brethouwer2007} delineating parameter regimes when $Pr \gtrsim 1$, these regime boundaries become $Re_b  \gg Pr$ and $Re_b  \ll Pr$ respectively. Thus we observe that at low $Pr$, the stratified turbulent regime can be realised even if $Re_b$ is very small.

Greater precision on these regime boundaries, permitting the identification of the domain of validity of the stratified intermittent regime, can be determined from figure \ref{fig:dnsscalings}(b). If we assume that, for each Reynolds number, the transition between the stratified turbulent and stratified intermittent regimes occurs when $\eta \simeq 0.4$, then the boundary is given by $BPe \simeq 0.0016Re^2$. This provides the more precise condition for the stratified turbulent regime:
\begin{equation} \label{eqn:cond2precise}
1 \ll BPe \ll 0.0016Re^2,
\end{equation}
labelled in figure \ref{fig:dnsparameterspace}. We note that this regime does not intersect the region of linear stability, indicating that for certain Reynolds numbers for which instability occurs ($2^{1/4} < Re < 25$) this particular type of stratified turbulence does not exist.

From figure \ref{fig:dnsscalings}(b) we can also estimate that the transition between the stratified intermittent regime and the stratified viscous regime approximately occurs when $\eta \simeq 0.05$ irrespective of the Reynolds number, which would imply that the boundary is given by $BPe \simeq 4.6Re^2$. Thus a more precise condition for the stratified viscous regime is given by 
\begin{equation} \label{eqn:cond4precise}
BPe \gg 4.6Re^2; \quad Re > 2^{1/4}.
\end{equation}
For each Reynolds number, the stratified intermittent regime exists for intermediate values of $BPe$ between conditions (\ref{eqn:cond2precise}) and (\ref{eqn:cond4precise}). When combined with the converse of the condition for the unstratified regime (i.e. $BPe \gg 1$) and the condition for linear instabililty ($Re>2^{1/4}$), this regime condition becomes
\begin{equation}
\max \{1,0.0016Re^2 \} \ll BPe \ll 4.6Re^2; \quad Re>2^{1/4}.
\end{equation}
Figure \ref{fig:dnsparameterspace} shows that the stratified intermittent regime can exist for any value of $Re$, provided  that the system is linearly unstable.


\section{Discussion}
\label{sec:discuss}

As summarized in section \ref{sec:regimes}, our numerical experiments have revealed that stratified horizontal Kolmogorov flows at high Reynolds number but low P\'eclet number exhibit (at least) four different non-trivial dynamical regimes depending on the respective values of the parameters $BPe$ and $Re$ (where $B$, $Pe$ and $Re$ were defined in (\ref{eq:BRedef}) and (\ref{eq:Pedef})). In all but one of these regimes, well-defined dominant balances in the momentum equation lead to simple scaling laws for the turbulent properties of the flow. We now first compare our results with prior studies of stratified mixing in the geophysical context, and then discuss the
 implications of our findings for stratified mixing in stars, whose understanding motivated this study.

\subsection{Comparison with stratified mixing in geophysical flows}

As we have demonstrated in this work, geophysical and astrophysical stratified turbulence is fundamentally different, because the former has a Prandtl number $Pr \gtrsim O(1)$ while the latter has $Pr \ll 1$. Therefore, crucially, in geophysically-relevant flows, a high Reynolds number flow necessarily also has a high P\'eclet number. Meanwhile, in astrophysics it is possible to have both $Re \gg 1$ and $Pe \ll 1$, and the effect of thermal diffusion can become a dominant factor in the system dynamics. As demonstrated by \cite{Lignieres1999} \citep[see also][]{Spiegel1962,Thual1992}, temperature and velocity fluctuations in the low P\'eclet number limit are slaved to one another, and density layering is prohibited (see section \ref{sec:intro2}). 
This is in stark contrast with geophysical flows where density layering (or at the very least, the propensity to form alternating regions of shallower and steeper density gradients) is key to understanding the properties of stratified turbulence in the LAST regime. Indeed, the standard Miles-Howard stability criterion \citep{Miles61,Howard61} for linear instability to vertical shear, namely $Ri_g < \frac{1}{4}$ (where $Ri_g$ is here the minimum gradient Richardson number based on the local vertical stratification and vertical shear), is at first glance incompatible with the ubiquitous presence of turbulence in  most large-scale stratified shear flows in geophysics (in particular in the ocean and atmosphere) where the gradient Richardson number is typically much larger than one, or indeed is irrelevant in the case of horizontally-sheared flows. However, small-scale layering releases this constraint by creating regions where the stratification is locally reduced, and the instability that is now allowed to develop continues to mix the layer, thereby allowing turbulence to sustain itself. This process, as reviewed in section \ref{sec:intro1}, can lead to the formation of layers on the scale $U_c / N_c$, and is controlled by the buoyancy Reynolds number $Re_b = Re Fr^2 = Re/B$. 

In astrophysics, typical values of the gradient Richardson number are also very large, but density layering is prohibited so this pathway to turbulence is not available. Instead, we have shown that three-dimensional perturbations of the horizontal shear \citep[see also][]{Deloncle2007,AroboneSarkar2012,Lucas2017} cause the flow to develop layers in the velocity field that enhance the vertical shear (or create it when it is not initially present). For sufficiently thin velocity layers, thermal diffusion reduces the effect of stratification, allowing vertical  shear instabilities to develop in between the layers. These two effects combine to drive turbulence and can cause substantial vertical mixing even when the background flow has no vertical shear. 
The dynamics of the system are no longer controlled by $Re Fr^2$, but instead, first by $BPe = Pe/Fr^2$ in the limit where  $BPe \ll Re^2$, and then by the ratio $BPe / Re^2$ in the limit where $BPe \gg Re^2$, thus partitioning parameter space in the four different dynamical regimes discussed in section \ref{sec:nonlinearregimes}.

The viscous regime that we have identified (when $BPe \gg Re^2$) is analogous to the viscously affected $Re_b \lesssim O(1)$ regime discussed by \cite{Brethouwer2007}, in the sense that it relies on the same dominant balances in the momentum equation. As a result, it exhibits the same scaling in terms of the vertical length scale $l_z \sim Re^{-1/2}$. It differs, however, in the treatment of the buoyancy equation, which is not surprising given the low P\'eclet number limit appropriate in our case. On the other hand, the stratified turbulent regime identified here bears little resemblance with the $Re_b \gg 1$, high $Pe$ regime of \cite{Brethouwer2007} (i.e. the LAST regime), where $l_z \sim U_c /N_c$. Indeed, for this new low P\'eclet number stratified turbulent regime,   
\begin{equation} \label{eqn:OurLengthScale}
l_z \sim (BPe)^{-1/3} L_c \sim \left( \frac{U_c \kappa}{N_c^2 } \right)^{1/3} 
\end{equation}
as found in section \ref{sec:nonlinearregimes}. From a dimensional analysis point of view, this new scaling can be understood  as the only combination of $U_c$ and $N_c$ that can be created to form a length scale given the constraint that $N_c^2$ and $\kappa$ can only appear together as $N_c^2/\kappa$ in the low P\'eclet number limit (as is apparent from (\ref{eq:lpeq})). 
But more importantly, we also saw that this scaling emerges from the assumption that the turbulent eddies are isotropic on the small scales, with $l_x \sim l_y \sim l_z$ (see section \ref{sec:stratreg}), which is quite different from the inherently anisotropic scalings discussed in \cite{Brethouwer2007} where $l_x, l_y \gg l_z$.  In other words, the stratified turbulent regime identified here is (we believe) a genuinely new regime of turbulence, that can only exist at low P\'eclet number, and so we refer to it as low P\'eclet number stratified turbulence, LPNST.

\subsection{Implications for mixing in stars}

We begin by comparing our numerical results to the theory proposed by \cite{Zahn92} for turbulence driven by horizontal shear in stellar radiation zones. Recall (see section \ref{sec:intro2}) that the characteristic flow length scale and amplitude in his model are given by (\ref{eq:horizscalings}). Written in terms of the non-dimensionalization used in this work (see section \ref{sec:model}), these are
\begin{equation}
L_c \sim \left( \frac{\epsilon^{1/3} }{BPe } \right)^{3/8} \quad \mbox{ and } \quad U_c \sim  \left( \frac{\epsilon^3 }{BPe } \right)^{1/8} ,
 \end{equation}
where $\epsilon$ is the non-dimensional dissipation rate \citep[see also][]{Lignieres2018}. As he assumes that all the energy input in the system (i.e. $\langle u \sin (y) \rangle$, which is always of order one in the chosen units) is dissipated viscously (i.e. there is negligible irreversible conversion into the potential energy reservoir), then $\epsilon \simeq 1$. The corresponding non-dimensional turbulent diffusivity in Zahn's model would therefore scale as
\begin{equation}
D_{\rm turb} \sim \left(BPe\right)^{-1/2},
\end{equation}
which is indeed what we find in the stratified turbulent regime, see (\ref{eq:strat_turb_diff}). It is interesting to note, however, that $L_c \sim (BPe)^{-3/8}$ in Zahn's model, and that this does not fit the data as well as our proposed $l_z \sim (BPe)^{-1/3}$ scaling. It is our belief that both scalings are relevant, with the difference emerging due to different choices for the velocity $U_c$. In (\ref{eqn:OurLengthScale}), we have assumed that $U_c$ is the r.m.s. horizontal velocity which is approximately constant in our simulations. By contrast, Zahn assumes a constant dissipation rate $\varepsilon$ in (\ref{eqn:ZahnLengthScale}), giving a modified Ozmidov scale representing the scale below which an isotropic turbulent cascade can exist \citep[see][]{Lignieres2018}.


While we believe our results are a step forward in the study of stratified mixing in stars, they are nevertheless not yet applicable {\it as is} for a number of reasons. First and foremost is the fact that the majority of stars (i.e. all stars except the most massive ones, see section \ref{sec:intro2}) are actually in the \emph{high P\'eclet} number yet low Prandtl number regime, while the simulations presented here only probe the low P\'eclet number regime. 
Indeed, a classic example of a stellar shear layer is the solar tachocline. Located just below the base of the solar convective envelope \citep{JCDSchou1988,Goode1991}, this layer contains a horizontal shear flow with characteristic values of the amplitude and length scale of the base flow being $U_c \simeq 150$ m/s and $L_c \simeq 5\times 10^8$m, while the buoyancy frequency is of the order of $N_c \simeq 10^{-3} \mathrm{s}^{-1}$ \citep{Tachobook2007}. With $\nu \simeq 0.001$m$^2$/s and $\kappa \simeq 1000$m$^2$/s, this implies $Re \sim O(10^{14})$, $Pe \sim O(10^8)$, and $B \sim O(10^7)$, with $Pr \sim O(10^{-6})$. Corresponding numbers for other main sequence low-mass and intermediate mass stars are in the same parameter regime. Our low P\'eclet number findings are not to be casually dismissed, however. As shown by \citet{Garaud2020}, flows with $Pe \gg 1$ and $Pr \ll 1$ can still be governed by low P\'eclet number dynamics (and therefore all the scalings derived in this work) when the turbulent P\'eclet number $Pe_t = w_{rms} l_z Pe \ll 1$. This is likely because the effective local P\'eclet number of the flow (written in terms of the actual vertical eddy scale $l_z$ instead of $L_c$) is low even though the P\'eclet number based on the global scale itself is large. A thorough study of the high P\'eclet and low Prandtl number regime is beyond the scope of this paper, however, and will be the subject of future work.

More crucially, however, is the fact that other effects will need to be taken into account before a comprehensive model of stratified mixing in stars can be created. The main source of shear in stars is their differential rotation, where the mean rotation rate is typically substantially larger than the shearing rate, and where the horizontal shear is usually global (i.e. with a length scale of the order of the stellar radius). This implies that  the effects of curvature and angular momentum conservation must be taken into account to determine whether the horizontal shear is unstable in the first place. Two-dimensional horizontal shear flows in a rotating spherical shell were first studied by \cite{Watson1980} \citep[see also][]{Garaud2001}, who found that the shearing rate must exceed a critical threshold for instability to proceed. In the context of our work, this implies that rotation could in principle inhibit the development of the primary instability. If the latter does take place, however, we anticipate that the same sequence of instabilities resulting in the development of small-scale eddies of size $l_z$ would ensue. The Rossby number based on $l_z$ is likely very large (in the tachocline, for instance, Ro $\sim U_c / \Omega l_z \sim O(10^4)$, where $\Omega \sim 3 \times 10^{-6}$s$^{-1}$ is the mean rotation rate of the Sun), suggesting that rotation would not have a significant effect on the flow dynamics in any stratified turbulent regime. It may be relevant in the intermittent and viscous regimes on the other hand, where the horizontal eddy scale is of the order of the scale of the background flow.

In addition, stars are subject to vertical shear as well as horizontal shear, and the dynamics of shear-induced turbulence are notably different in the two cases (see section \ref{sec:intro2}). A question of interest will therefore be to establish what controls the outcome when vertical and horizontal shear are both present. Finally, most stars are expected to be magnetized to some extent \citep{mestel2012stellar}, either by the presence of a primordial magnetic field or by the action of a dynamo in a nearby convective zone. The effect of these magnetic fields will need to be taken into account to construct a  truly astrophysically relevant theory of stratified turbulence. \\



This work was initiated as a project at the Woods Hole Geophysical Fluid Dynamics summer program in 2018. The authors thank the program for giving them the opportunity to collaborate on this topic and for their financial support. L.C. was also supported by the Natural Environment Research Council (grant number NE/L002507/1). P.G. was also funded by NSF AST 1517927. The research activity of C.P.C. was supported  by the EPSRC Programme Grant EP/K034529/1 entitled ‘Mathematical Underpinnings of Stratified Turbulence’. All simulations presented here were performed on either the Hyades computer, purchased at UCSC with an NSF MRI grant, or the NSF XSEDE supercomputing facilities (Comet). The authors thank S. Stellmach for providing his code, and F. Ligni{\`e}res for crucial comments that fixed an incorrect statement in the original manuscript. The data used to create figure \ref{fig:dnsscalings} are available at https://doi.org/10.17863/CAM.54365.


\section*{Declaration of Interests}

\noindent
The authors report no conflict of interest.


\appendix
\section{Generalisation of the linear stability analysis}\label{appA}


It is sometimes of interest, particularly when comparing results with DNSs, to consider the linear stability of a laminar flow with different amplitude to that of the basic laminar solution (\ref{eqn:laminarsolution}). Consequently, we explain here how the linear stability of such flows can be computed from the results presented in section \ref{sec:linearstability}. We will focus on the standard system of equations (\ref{eqn:momentumnondim})-(\ref{eqn:continuitynondim}), although an equivalent procedure can also be applied straightforwardly to the LPN equations (\ref{eqn:lpnequation1})-(\ref{eqn:lpnequation2}). 

We consider a laminar flow $a \mathbf{u}_L(y)$ given by
\begin{equation}
a \mathbf{u}_L(y) = aRe \sin (y) \mathbf{e}_x,
\end{equation}
with amplitude $a Re$ where $a \in \mathbb{R}$, and without loss of generality $a> 0$. For small perturbations $\mathbf{u}'(x,y,z,t)$ away from this laminar flow, i.e. letting $\mathbf{u} = a \mathbf{u}_L(y) + \mathbf{u}'(x,y,z,t)$, and using the assumption that the growth rates of instabilities are significantly larger than the rate at which the background flow is evolving due to the uncompensated forcing, the linearised perturbation equations are:
\begin{equation}\label{eqn:linear1b}
\frac{\partial \mathbf{u}'}{\partial t} + a Re   \cos (y) v' \mathbf{e}_x + a Re \sin(y) \frac{\partial \mathbf{u}'}{\partial x} + \nabla p = \frac{1}{Re} \nabla^2 \mathbf{u}' + BT' \mathbf{e}_z,
\end{equation}
\begin{equation}\label{eqn:linear2b}
\frac{\partial T'}{\partial t} + a Re \sin(y) \frac{\partial T'}{\partial x} +  w' = \frac{1}{RePr} \nabla^2 T',
\end{equation}
\begin{equation}\label{eqn:linear3b}
\nabla \cdot \mathbf{u}' = 0.
\end{equation}
Rescaling the velocity field according to $\mathbf{u}' = a^{1/2} \mathbf{\tilde{u}}'$ leads to the transformed set:
\begin{equation}\label{eqn:linear1a}
a^{-1/2}\frac{\partial \mathbf{\tilde{u}}'}{\partial t} + a^{1/2} Re   \cos (y) \tilde{v}' \mathbf{e}_x + a^{1/2}Re \sin(y) \frac{\partial \mathbf{\tilde{u}}'}{\partial x} + a^{-1} \nabla p = \frac{1}{a^{1/2}Re} \nabla^2 \mathbf{\tilde{u}}' + \frac{B}{a}T' \mathbf{e}_z,
\end{equation}
\begin{equation}\label{eqn:linear2a}
a^{-1/2}\frac{\partial T'}{\partial t} + a^{1/2} Re \sin(y) \frac{\partial T'}{\partial x} +  \tilde{w}' = \frac{1}{a^{1/2}RePr} \nabla^2 T',
\end{equation}
\begin{equation}\label{eqn:linear3a}
\nabla \cdot \mathbf{\tilde{u}}' = 0.
\end{equation}
By considering normal mode disturbances of the form $q(x,y,z,t) = \hat{q}(y) \exp[ik_x x + ik_z z + \sigma t]$ and rescaling the parameters and growth rates using the relations:
\begin{equation} \label{eqn:linear_transformation}
\hat{\sigma} = \frac{\sigma}{a^{1/2}}; \quad \hat{Re}=a^{1/2}Re; \quad \hat{B} = \frac{B}{a}; \quad \hat{Pr}=Pr,
\end{equation}
the resulting system is identical to the set of equations (\ref{eqn:lsstandard1})-(\ref{eqn:lsstandard5}) except for the rescaling implicit in the hats on parameters and growth rates. It can be re-formulated as a generalised eigenvalue problem for the complex growth rates $\hat{\sigma}$,
\begin{equation} \label{eqn:linearstabA}
\mathbf{A}(k_x,k_z,\hat{Re},\hat{B},\hat{Pr}) \mathbf{X} = \hat{\sigma} \mathbf{B}\mathbf{X},
\end{equation}
and can solved using the method described in section \ref{sec:linearstability}.

The linear stability analysis presented in section \ref{sec:linearstability} considered $a=1$, where $\hat{Re}=Re$, $\hat{B}=B$, $\hat{Pr}=Pr$ and $\hat{\sigma}=\sigma$. For $a \ne 1$, relations (\ref{eqn:linear_transformation}) provide a transformation between the original analysis and the linear stability of flows with generic amplitude $a \mathbf{u}_L(y)$.

\section{Critical Reynolds number for linear instability} \label{appB}

An important finding in this study, determined numerically across a broad spectrum of parameters, is the fact that the critical Reynolds number, $Re_c$, for the onset of linear instability, as given by the 2D mode ($k_z=0$), is independent of both the stratification and Prandtl numbers, being fixed at $Re_c=2^{1/4}$. This result holds for both the standard equations and the LPN equations and differs quite substantially from that obtained in \citet{Garaudal15} for the case of a vertically orientated shear, where stratification was found to be able to stabilize a system.

To see why this is the case, we observe in equations (\ref{eqn:lsstandard1})-(\ref{eqn:lsstandard5}) (or the equivalent LPN equations (\ref{eqn:lslpn1})-(\ref{eqn:lslpn4})) that setting $k_z=0$ reduces the problem to the study of equations (\ref{eqn:lsstandard1}), (\ref{eqn:lsstandard2}) and (\ref{eqn:lsstandard5}) (or equivalently (\ref{eqn:lslpn1}), (\ref{eqn:lslpn2}) and (\ref{eqn:lslpn4})). This reduced problem is well studied \citep[]{Beaumont1981,Balmforthyoung}, being the linear stability of an unstratified ($B=0$) flow. The critical Reynolds number for instability around a basic state of $\mathbf{u}_L(y) = \sin (y) \mathbf{e}_x$ has been shown to be $\sqrt{2}$ \citep{Beaumont1981}. As detailed in appendix \ref{appA}, a simple transformation given by relations (\ref{eqn:linear_transformation}) (for $a=Re^{-1}$) gives the corresponding critical Reynolds number for 2D modes in this study to be $2^{1/4}$, which corresponds to the result obtained numerically. Consequently, we note that horizontally-sheared Kolmogorov flows with $Re>2^{1/4}$ are always unstable, irrespective of the stratification, and thus form a convenient basis from which to study the subsequent nonlinear evolution of stratified, low-P\'{e}clet number flows.

\bibliography{stratified_horizontal_shear}
\bibliographystyle{jfm}

\end{document}